\documentclass[aps,pra,reprint,groupedaddress,amsmath,amssymb]{revtex4-1}
\pdfoutput=1

\usepackage{hyperref}
\newcommand{\rmi}{\mathrm{i}}

\DeclareMathOperator{\tr}{tr}

 \newcommand{\vect}[1]{\mathbf{#1}}

\newcommand{\ket}[1]{\ensuremath{\lvert #1 \rangle}}

\newcommand{\outprod}[2]{\ensuremath{\lvert #1 \rangle \langle #2 \rvert}}

\newcommand{\abs}[1]{\lvert #1 \rvert}
\newcommand{\norm}[1]{\lVert #1 \rVert}

\newcommand{\eqcl}[1]{\lbrack #1 \rbrack}

\begin{document}

\title{Emergence of outcomes in quantum mechanics}

\author{Bradley A. Foreman}
\affiliation{Department of Physics, The Hong Kong University of Science and 
Technology, Clear Water Bay, Kowloon, Hong Kong, China}

%\date{\today}

\begin{abstract}
A persistent focus on the concept of emergence as a core element of the
scientific method allows a clean separation, insofar as this is possible, of the
physical and philosophical aspects of the problem of outcomes in quantum
mechanics. The philosophical part of the problem is to explain why a closed
system has definite experimental outcomes. The physical part is to show
mathematically that there exists a limit in which the contradiction between
unitary Schr\"odinger dynamics and a reduction process leading to distinct
outcomes becomes negligible according to an explicitly stated criterion, and to
make this criterion as objective as possible. The physical problem is solved
here by redefining the notion of a quantum state and finding a suitable measure
for the change of state upon reduction. The appropriate definition of the
quantum state is not as a ray or density operator in Hilbert space, but rather
as an equivalence class consisting of all density operators in a given subspace,
the members of which all describe the same experimental outcome. For systems
containing only subsystems that are integrated with their environments, these
equivalence classes can be represented mathematically by projection operators,
and the resulting formalism is closely related to that used by von Neumann to
study the increase of entropy predicted by the second law of thermodynamics.
However, nearly isolated subsystems are reduced only indirectly, as a
consequence of their interaction with integrated subsystems. The reduced states
of isolated subsystems are the same conditional states used in the definition of
quantum discord. The key concepts of decoherence theory can all be adapted to
fit this definition of a quantum state, resulting in a unified theory capable of
resolving, in principle, all aspects of the quantum measurement problem. The
theory thus obtained is weakly objective but not strongly objective.
\end{abstract}

%\keywords{}

\maketitle

\section{Introduction}

The measurement problem has been the central mystery of quantum mechanics for
more than 90 years \cite{[] [{; English translation: }] vonNeumann1932,
*vonNeumann1955, Wigner1963, WheelerZurek1983, Bell2004, dEspagnat1976,
Leggett2005, Laloe2019, Weinberg2015n}. It comprises three distinct but
interrelated issues \cite{Schlosshauer2007}. The \emph{interference problem} is
to explain why quantum interference effects are difficult to observe, especially
on macroscopic scales. The \emph{preferred-basis problem} is to define which
particular set of variables is actually observed. The \emph{problem of
outcomes}, also called the and--or problem \cite{Schrodinger1933,
JauchWignerYanase1967, LevyLeblond1977, Bell1990, dEspagnat1995}, is to
reconcile our perception of a unique experimental outcome with the superposition
of many outcomes generated by the Schr\"odinger equation. How is it physically
possible to perceive Schr\"odinger's cat
\cite{Schrodinger1935a_pt123, *Trimmer1980}
%\cite{WheelerZurek1983, Laloe2019} 
as dead \emph{or} alive, rather than as a superposition of dead \emph{and} alive
(whatever that might mean)? The ultimate goal is to reconcile the notorious
``reduction'' or ``collapse'' of the quantum state with the unitary dynamics of
a closed system.

It is well known that this cannot be done in the sense of a strict logical
deduction, because the two processes are fundamentally contradictory
\cite{vonNeumann1955, Wigner1963, dEspagnat1976, Fine1970, BassiGhirardi2000}.
The goal is rather to show that reduction \emph{emerges} from unitary dynamics,
in much the same way that temperature in thermodynamics emerges from statistical
mechanics. That is, we must show that there exists some \emph{limit} in which
the discrepancy between the two processes vanishes. This technique is ubiquitous
in the physical sciences \cite{Anderson1972, Primas1983, Primas1998, Berry1994,
Berry2002, LaughlinPines2000, Laughlin2005, Schweber1993, Batterman2002,
Bokulich2008, Chibbaro2014}; from the standard model of quantum field theory
\cite{[] [{, p.\ 499.}] WeinbergVol1} on up, there is no example of an
empirically validated scientific theory that is not widely accepted to be
emergent.

It is now generally acknowledged that, in this sense, the theory of decoherence
\cite{Schlosshauer2007, JoosZeh2003, Zurek1991, Zurek2003, HarocheRaimond2006}
has solved both the interference problem and the preferred-basis problem
\cite{Schlosshauer2007}. The problem of outcomes, however, remains unsolved
\cite{Schlosshauer2007, JoosZeh2003, Adler2003}. This is a problem in physics,
not in philosophy or interpretation. Faced with an open contradiction at the
heart of a fundamental physical theory, the job of the physicist \cite{[] [{;
quotations from p.\ 471.}] Feynman1982} is either to ``define the real problem''
by showing that the theoretical conflict has definite experimental implications,
or to prove ``that there's no real problem'' by showing that the contradiction
can be resolved as a case of emergence. Feeling ``nervous'' at our sustained
inability to finish this job is the mark of a good scientist \cite{Feynman1982},
not of a bad philosopher or interpreter.

Here I show that the problem of outcomes can indeed be resolved as a case of
emergence. A crucial element of the solution is finding a suitable measure for
the change in the quantum state upon reduction. The solution is not trivial,
because it only works if the quantum state itself is redefined. The quantum
state should be defined neither as a ray in Hilbert space nor as a statistical
mixture of such rays (i.e., as a density operator), but as an \emph{equivalence
class} \cite{vonNeumann1955secV4, Jauch1964, [] [{, Chap.\ 11.}] Jauch1968} of
density operators. The properties of a closely related equivalence class were
studied long ago by von Neumann \cite{vonNeumann1955secV4}, but his results have
since been all but forgotten. The central message of this paper is that a
modified version of von Neumann's definition allows the last remaining part of
the quantum measurement problem to be solved. This modification unifies von
Neumann's theory with various elements of modern decoherence theory, including
the concept of a conditional state used in the definition of quantum discord
\cite{Zurek2000, OllivierZurek2001, HendersonVedral2001, ModiVedral2012}. A
brief account of the equivalence-class formulation of the quantum state is
presented elsewhere \cite{Foreman2019b}, focusing on the central role played by
Bohr's concept of complementarity.

It should be stressed that this emergence-based solution does not answer
philosophical questions such as why measurements have outcomes at all, or what
it is that selects a particular outcome \cite{Schlosshauer2007}. However, as
discussed below, it is doubtful whether pilot-wave \cite{BohmHiley1993,
Holland1993, DurrGoldsteinZanghi2013} or dynamical-reduction
\cite{BassiGhirardi2003, BassiLochan2013} theories truly answer these questions
either, because their ontological claims are based on concepts that seem to be
entirely explicable as emergent.

The paper begins in Sec.\ \ref{sec:emergence_meaning} with a discussion of how
the concept of emergence allows us to reconcile the often contradictory demands
of different levels of physical theory. Section\ \ref{sec:science_contradictory}
argues that such contradictions are inherent in the scientific method itself.
Section \ref{sec:reduction_process} reviews the mathematical formalism for the
reduction process in conventional quantum mechanics. Section
\ref{sec:emergent_noninterference} reviews the use of decoherence theory to
justify the emergence of noninterference between conditional probabilities in a
quantum history. The significance of various possible metrics for the emergence
of outcomes is studied in Sec.\ \ref{sec:testing_outcomes}.

Section \ref{sec:emergence_outcomes} shows that the problem of outcomes can be
resolved in the special case of integrated subsystems by von Neumann's
definition of equivalence classes. Comparison between theory and experiment is
made possible by the emergence of statistical frequencies in an ensemble, as
described in Sec.\ \ref{sec:emergent_statistics}. The theory is extended to
include isolated subsystems in Sec.\ \ref{sec:partial_reduction}. The
preferred-basis problem is addressed in Sec.\ \ref{sec:preferred_basis} using a
modified version of Zurek's predictability sieve. Section
\ref{sec:weak_objectivity} discusses the weakly objective nature of the
solutions and examines various claims of strong objectivity in quantum
mechanics. Reasons why the concept of emergence is applicable to the quantum
measurement problem are further discussed in Sec.\ \ref{sec:emergent_reality}.
Finally, Sec.\ \ref{sec:conclusions} summarizes the conclusions of the paper.

\section{What emergence means}

\label{sec:emergence_meaning}

It is important to clarify at the outset some possible misconceptions
\cite{[{Section \ref{sec:emergence_meaning} should not be misconstrued as
suggesting that Anderson shares any of these possible misconceptions; this is
clearly not the case, as the following quotation shows: ``The important lessons
to be drawn are two: 1) totally new physics can \emph{emerge} when systems get
large enough to break the symmetries of the underlying laws; 2) by construction,
if you like, those \emph{emergent properties} can be completely unexpected and
intellectually independent of the underlying laws, and have no referent in
them.'' }] [{, p.\ 144. As stressed in Sec.\ \ref{sec:emergence_meaning}, this
intellectual independence is not limited to obvious cases such as the emergence
of universal behavior at critical points described by renormalization-group
methods; it is relevant also in the more mundane case of ordinary
thermodynamics.}] Anderson2011} that could arise from popular slogans such as
``more is different'' \cite{Anderson1972} and from the idea that the elementary
entities of an emergent level of description ``obey the laws''
\cite{Anderson1972} of the more fundamental level. These might give the
impression that the concept of emergence is inapplicable to such fundamentally
disparate processes as quantum-state reduction and unitary dynamics. However,
this is not so; there is no essential difference between this case of emergence
and that of, say, temperature in thermodynamics.

Thermodynamics emerges from statistical mechanics in the so-called thermodynamic
limit $N \to \infty$, $V \to \infty$, $N/V =$ constant, in which $N$ is the
number of particles in a system and $V$ is its volume. Of course, in any real
experiment, $N$ and $V$ are finite, but typically $N \gg 1$. Invoking the
thermodynamic limit is a signal that one intends to ignore small terms involving
$\delta = 1 / N$ in certain parts of the theory, while usually retaining them in
other parts of the theory. The validity of the limit simply means that if
neglecting these terms gives acceptable results when $\delta = \delta_0 \ll 1$,
then the same will be true for all other cases in which $\delta < \delta_0$;
there is no need to investigate these cases individually.

Ignoring $\delta$ in certain parts of a theory means setting $\delta = 0$ when
actually $\delta > 0$. This changes the logical structure of the theory, so that
many things that were true in the original theory become false in the modified
theory, and vice versa. The two theories are thus logically contradictory for
all cases of practical relevance (i.e., when $\delta > 0$). Within the modified
logical framework, there may be some new concepts that are immensely powerful in
elucidating the structure of the new theory (i.e., they vastly increase its
algorithmic compressibility). These are referred to as emergent concepts.
Temperature and entropy are examples of such concepts in thermodynamics.

These quantities can already be defined in statistical mechanics by means of the
canonical or microcanonical probability distribution. Other thermodynamic
variables are limited to a few statistical moments of the probability
distribution, such as mean values and correlation functions, but this
restriction is not the primary issue of concern. The problem is that temperature
and entropy are defined operationally in thermodynamics via the properties of
interacting subsystems. For example, temperature is defined operationally as
what is measured by a thermometer, which is an open system that functions by
interacting with its local environment for a long time (relative to the
relaxation time) until they come to equilibrium. Of course, equilibrium is
itself an emergent concept that never holds exactly in any real experiment.

In order to apply the statistical-mechanical definitions of temperature and
entropy to such a situation, we must introduce subsystems
\cite{LandauLifshitz1980a, Lenard1978} and separate the energy of a closed
system into several parts. The \emph{external} energy includes the bulk kinetic
energy of each subsystem and the macroscopic energy of interaction between the
subsystems, such as the gravitational energy of interaction between small volume
elements in a planet or star. The \emph{internal} energy of each subsystem
depends on its internal microscopic properties, the details of which are
regarded as irrelevant in thermodynamics. The temperature and entropy of each
subsystem are defined in terms of its internal energy. Both external and
internal energy are thermodynamic variables.

What remains is the \emph{microscopic} energy of interaction between subsystems.
This is not a thermodynamic variable. It is ignored in thermodynamics even
though it is nonzero in statistical mechanics. The justification for this is
that all long-range interactions are assumed to have been included in the
macroscopic energy of interaction. Thus, the microscopic energy of interaction
may be regarded as a short-range surface effect, which becomes negligible in
comparison to the internal energy of the subsystems (a volume effect) for
regularly shaped subsystems in the thermodynamic limit. One could conceivably
also argue that, on average, this energy should also vanish (or at least reach a
minimum) in the limit of thermal equilibrium, independent of the thermodynamic
limit. In either case, ignoring this energy makes it possible to formulate
conservation of energy in terms of thermodynamic variables alone.

But this shows that the law of conservation of energy in thermodynamics
contradicts the law of conservation of energy in statistical mechanics. These
two laws can be reconciled only if we agree to ignore the microscopic energy of
interaction. Of course, this approximation cannot be invoked outside this narrow
context. The microscopic energy of interaction is essential, for instance, if we
wish to describe the dynamics of nonequilibrium systems---it gives rise to the
existence of heat transfers and the attainment of thermal equilibrium, both of
which are essential to the empirical definitions of entropy and temperature.

The link thus established between statistical mechanics and thermodynamics is
not trivial, due to the long range of the Coulomb interaction. One must define
subsystems carefully so that all Coulomb terms can be clearly separated
\cite{BornHuang1954} into (1) long-range macroscopic effects that appear only in
the external energy and (2) screened short-range effective interactions that
appear in both the internal energy and the microscopic energy of interaction.
The emergence of temperature in thermodynamics therefore requires, in addition
to the thermodynamic limit, another level of emergence in which well-defined
subsystems with these properties emerge from a collection of charged particles.

The notion of a state in thermodynamics is thus fundamentally different from
that in statistical mechanics, because the thermodynamic state has the presumed
existence of subsystems with negligible microscopic energy of interaction built
into its very foundations \cite{LandauLifshitz1980a, Lenard1978}---otherwise it
is experimentally vacuous. This mandatory change in the definition of state is a
key indicator that the thermodynamic temperature and entropy are truly emergent
quantities. The essence of this constraint was already well understood by Gibbs
in 1902 \cite{[] [{, p.\ 37.}] Gibbs1902}.

Whether subsystems with the requisite properties can be defined also depends
strongly on which phases of matter are present in the system.  Different phases
must be described by different thermodynamic variables, so the choice of
thermodynamic variables is also connected with the definition of subsystems.

Hence, this example shows clearly that thermodynamics does not ``obey the laws''
of statistical mechanics in any case of practical relevance (i.e., for finite
$N$). The two theories may give very similar predictions if the conditions are
right, but they remain logically contradictory. The contradiction can be
tolerated only if we choose to ignore certain small but nonzero quantities. It
is this choice that makes the difference, not the mere presence of ``more.''

The conflict between unitary dynamics and quantum-state reduction is no greater
than that between the two laws of energy conservation in statistical mechanics
and thermodynamics. In both cases the emergent theory contradicts basic
properties of the fundamental theory. The conflict is resolved by identifying a
suitable limit in which the contradiction can be ignored.

\section{Science is contradictory}

\label{sec:science_contradictory}

\begin{quote}
Secret, secret, close the doors! \\
\phantom{x} \hfill --- R. P. Feynman \cite{Feynman1982}
\end{quote}

Many physicists are uneasy with the idea of drawing an analogy between
thermodynamics and quantum mechanics. Bell, for example, has eloquently
expressed the opinion \cite{Bell2004} that quantum mechanics should be
formulated ``precisely,'' without reference to approximations or the concept of
emergence. Bell is more than happy to use approximations in practical
applications, but he feels they should play no role in the basic formulation of
any theory that has aspirations to be fundamental.

There is, however, a case to be made for the claim that the most important
characteristic of a fundamental theory is not ``precision'' but universality. A
fundamental scientific theory should permit the existence, at least in
principle, of all that we see around us, including scientists. In particular, it
should allow the evolution of ``information gathering and utilizing systems''
\cite{GellMannHartle1990} or protoscientists with the ability to profit from
even the barest rudiments of the scientific method. But this points away from
``precision'' and toward the inclusion of emergence at the most fundamental
level, because the scientific method is inherently self-contradictory.

The first requirement of science is that it should refer only to things that are
\emph{objective}, in the sense that they do not depend on the perspective of any
observer. In quantum mechanics, this level is described by the unbroken symmetry
of the quantum state vector, evolving unitarily in accordance with the
Schr\"odinger equation.

The second requirement of science is that it should be capable of describing
differences.  The most primitive scientific statement is the bare observation
that ``\emph{this} is different from \emph{that}.''  But this requires some
way of assigning meaning to \emph{this} and \emph{that}.  In quantum
mechanics, this is done by defining subsystems.  However, there is no
objective way to do this; subsystems are inherently contextual entities 
that always refer, at least implicitly, to the perspective of an observer.

The third requirement of science is that it should be capable of describing
regularities. The laws of science refer only to what is repeatable. But no two
things in this world are ever \emph{exactly} the same, if only because they
occur at different places or different times, and in different environments. A
protoscientist will not get very far if it is only capable of being perpetually
astounded by the newness of it all. Regularities come into existence only after
the protoscientist declares what is relevant and what is irrelevant. In quantum
mechanics, this is done in part by defining collective variables (to be
discussed further below), and in part by discarding irrelevant portions of the
quantum state during the reduction process. Regularities, the raw material of
science, are therefore unavoidably subjective.

Hence, the most basic principles of the scientific method demand that science be
simultaneously objective and subjective. The only way to combine these
conflicting requirements in a single theory is to veil the contradictions using
the technique of emergence. In this way the theory can
be made \emph{weakly} objective, in a sense to be defined below.

This is reminiscent of the story of the emperor's new clothes
\cite{Andersen1984}. The emperor (science) is forbidden to flaunt a naked
contradiction in public. Clothing the emperor in metaphysics will not do, for
there are many shrewd children in the audience; only mathematical garments
suffice. The fig leaf of emergence may not be much, but it is all that we have
and all that we need.

Various degrees of objectivity and subjectivity were defined by d'Espagnat
\cite{dEspagnat1976, dEspagnat1995}; the definitions adopted here are somewhat
modified \footnote{The most significant change is that d'Espagnat equates the
concepts of intersubjectivity and weak objectivity, whereas I define them to be
different.}. An overtly subjective theory is one that depends explicitly on the
capabilities of individual protoscientists. An intersubjective theory depends on
these capabilities, but all protoscientists are assumed to be roughly the same,
which makes it possible for them to reach mutual understanding and agreement. A
weakly objective theory is one that relies upon the concept of emergence, but
the quantities thereby ignored are far below the resolution limits of the
protoscientists, so the theory does not depend explicitly on the capabilities of
protoscientists. Thermodynamics is weakly objective in this sense, because the
coarse graining that is used to establish the second law of thermodynamics need
not be tailored explicitly for the limitations of human observers. Weak
objectivity strengthens intersubjective agreement by allowing protoscientists to
avoid (for the most part) squabbles over solipsism. But an element of
subjectivity always remains in a weakly objective theory.

Zeh has described the ideal limit in which the quantities declared to be
irrelevant do not refer directly to the knowledge of any actual observer as a
process of ``objectivization'' of a subjective theory \cite{Zeh2007}. Zurek has
stressed the importance of stability under environmental monitoring and the
redundancy of records in many subsystems for an operational definition of
objectivity \cite{Zurek1998, Zurek2007b}. Both of these are examples of weak
objectivity.

A strongly objective theory does not rely upon the concept of emergence as it is
usually understood. Of course, any scientific theory must declare some things to
be irrelevant in order to identify regularities, but in a strongly objective
theory it is assumed that meaningful regularities can be identified by declaring
\emph{arbitrarily small} quantities of information to be irrelevant, so that
nothing qualitatively new (e.g., quantum state reduction) need emerge. Bell
advocated strong objectivity as the target for all respectable formulations of
quantum mechanics \cite{Bell2004}.

However, it seems unlikely that protoscientists could ever evolve if their
information gathering is limited to arbitrarily weak regularities. There is not
much profit in defining a ``law'' for regularities that are arbitrarily few and
far between. To get the scientific enterprise off the ground, something more
seems to be required---something qualitatively new. This suggests that the
concept of emergence may be unavoidable---a conclusion supported by the above
observation that no examples of theories that are both empirically validated and
non-emergent are known to exist at present.

Weak objectivity is therefore taken to be the primary goal of the present
theory. Intersubjectivity would be allowed as a fallback position, but only if
weak objectivity is found to be unattainable.

\section{The reduction process}

\label{sec:reduction_process}

Let us now consider the mathematical formulation of the measurement 
problem.  The time evolution of the density operator $\rho$ of a closed
system is given by the Schr\"odinger equation as
\begin{equation}
\rho (t_b) = U (t_b, t_a) \rho (t_a) U (t_a, t_b) , \label{eq:unitary}
\end{equation}
in which $\rho (t)$ is the value of $\rho$ at time $t$, $U (t, t') = \exp[-\rmi
H (t - t') / \hbar]$ is the time evolution operator, $H$ is the Hamiltonian of
the system, and $\rho$ is assumed to be normalized ($\tr \rho = 1$).

\subsection{von Neumann reduction}

However, according to von Neumann \cite{vonNeumann1955}, this unitary evolution
is suspended during an ideal ``projective'' measurement of the operator
\begin{equation}
\Lambda = \sum_i \lambda_i P_i ,  \label{eq:observable}
\end{equation}
in which $P_i$ is a projection operator of rank one ($d_i \equiv \tr P_i = 1$).
These projectors are idempotent, hermitian, mutually orthogonal, and exhaustive:
\begin{equation}
P_i^2 = P_i = P_i^{\dagger} , \quad P_i P_j = \delta_{ij} P_i , \quad
\sum_i P_i = 1 . \label{eq:projector_criteria}
\end{equation}
The measurement effectively asks a set of yes--no
questions---namely, whether the state $\rho$ will be found to match the vector
subspace defined by $P_i$. At the time of the measurement, the unitary process
(\ref{eq:unitary}) is replaced by the two-stage reduction process
\begin{subequations}
\label{eq:reduction}
\begin{align}
\rho & \to \hat{\rho} = \sum_{i} w_i \rho_i ,
\label{eq:reduction_a} \\
\hat{\rho} & \to \rho_i , \label{eq:reduction_b}
\end{align}
\end{subequations}
in which
\begin{equation}
w_i = \tr (\rho P_i) \label{eq:weight}
\end{equation}
and
\begin{equation}
\rho_i = P_i \qquad (\tr P_i = 1) . \label{eq:rho_i_vN1}
\end{equation}
The first stage (\ref{eq:reduction_a}) accounts for the elimination of
interference, whereas the second stage (\ref{eq:reduction_b}) describes the
selection of an individual outcome. The probability of obtaining outcome
(\ref{eq:reduction_b}) is $w_i$.

\subsection{L\"uders reduction}

Von Neumann assumed that the measurement was maximally fine-grained in the sense
that $d_i = 1$. However, the outcome of the measurement is then ambiguous
whenever the spectrum of $\Lambda$ is degenerate (i.e., $\lambda_i = \lambda_j$
for $i \ne j$), because the corresponding projectors $P_i$ and $P_j$ are not
uniquely defined. To circumvent this problem, L\"uders \cite{[] [{; English
translation: }] Lueders1950, *Lueders2006} replaced the reduced state
(\ref{eq:rho_i_vN1}) with
\begin{equation}
\rho_i = \frac{P_i \rho P_i}{w_i} \qquad (w_i \ne 0) , \label{eq:rho_i_Lu}
\end{equation}
in which $d_i \ge 1$ is now allowed, but it is assumed that $\lambda_i \ne
\lambda_j$ for $i \ne j$. The probability of outcome (\ref{eq:rho_i_Lu}) is
still given by Eq.\ (\ref{eq:weight}). The L\"uders reduced state
(\ref{eq:rho_i_Lu}) minimizes the change $\rho \to \hat{\rho}$
%\cite{Lueders1950, GoldbergerWatson1964, BellNauenberg1966} 
associated with a given projector set
\begin{equation}
\mathcal{P} = \{ P_i \mid P_i^{\dagger} P_j = \delta_{ij} P_i , \sum_i P_i = 1
\} . \label{eq:P_set}
\end{equation}
The L\"uders reduction rule has since been adopted almost universally in
textbooks (e.g., \cite{Messiah1962, CohTan1977}) for the case of projective
measurements.

The L\"uders rule is sometimes attributed to von Neumann \cite{Wigner1963,
AharonovBergmannLebowitz1964}. However, von Neumann took a different approach to
this problem that is rooted in epistemology rather than conservation of $\rho$.
Since von Neumann's method \cite{vonNeumann1955secV4} is rarely used in modern
work on the measurement problem, discussion of his approach will be postponed to
Sec.\ \ref{sec:emergence_outcomes}.

\subsection{Closed systems}

In traditional presentations of the theory of measurement, reduction is applied
to an open system when it is ``measured'' from the outside by some other system
that is not included in the quantum-mechanical description. In this paper,
however, the theory is applied exclusively to \emph{closed} systems in which
measurement is just one type of interaction between subsystems. The projector
set $\mathcal{P}$ describes properties of subsystems that are of interest to some
protoscientist within the system, although the protoscientist may or may not be
included explicitly as a subsystem (or a set of subsystems) in the mathematical
definition of $\mathcal{P}$. The objective is to find a way to formulate the
reduction process as emergent, so that its conflict with the unitary dynamics
(\ref{eq:unitary}) can be neglected.

It is assumed throughout that the information accessible to the protoscientist
is limited to the outcomes (\ref{eq:rho_i_Lu}) and their associated
Born-rule probabilities (\ref{eq:weight}). However, at the outset, $\{ w_i \}$
is to be regarded as just a set of numbers derived from the mathematical
structure of Hilbert space. This set satisfies $w_i \ge 0$ and $\sum_i w_i = 1$
by definition, but the concept of probability will emerge only after further
theoretical developments. Nevertheless, in the interest of avoiding tortuous
linguistic constructions, the \emph{language} of probability theory will be used
henceforth, interspersed with occasional comments indicating the progress along
the path to emergence.

\subsection{Subsystem projectors}

The projector $P_i$ is henceforth defined as a product of subsystem
projectors:
\begin{equation}
P_i = \bigotimes_{k} \tilde{P}_{k i_k} = \prod_{k} P_{k i_k} ,  
\label{eq:compound_projector}
\end{equation}
in which $k$ labels the subsystems and $i_k$ is the $k$ component of $i = (i_1,
i_2, \ldots)$. $P_{k i_k}$ is an extension of the subsystem projector
$\tilde{P}_{k i_k}$ to the entire system:
\begin{equation}
P_{k i_k} = 1_1 \otimes \cdots 1_{k-1} \otimes \tilde{P}_{k i_k} \otimes 1_{k+1} 
\cdots \otimes 1_n , \label{eq:subsystem_projector0}
\end{equation}
in which $1_k$ is the identity operator for subsystem $k$ and 
the set $\{ \tilde{P}_{k i_k} \}$ satisfies
\begin{equation}
\tilde{P}_{k i_k}^{\dagger} \tilde{P}_{k j_k} = \delta_{i_k j_k} \tilde{P}_{k
i_k} , \qquad \sum_{i_k} \tilde{P}_{k i_k} = 1_{k} .
\label{eq:P_tilde_identities}
\end{equation}
The rank of $P_i$ is thus $d_i = \prod_k d_{k i_k}$, in which
\begin{equation}
d_{k i_k} = \tr \tilde{P}_{k i_k} \ne \tr P_{k i_k} .
\label{eq:subsystem_projector_dimensions}
\end{equation}
The dimension of the subspace corresponding to subsystem $k$ is
\begin{equation}
d_k = \sum_{i_k} d_{k i_k} = \tr 1_k . \label{eq:subsystem_dimension}
\end{equation}

\subsection{Microstates, macrostates, collective variables, and 
internal environments}

\label{sec:microstates_macrostates}

In a typical measurement situation, the time evolution (\ref{eq:unitary})
generates correlations between the \emph{microstates} of a subsystem $k$ and the
\emph{macrostates} of a measuring apparatus $k'$. Microstates and macrostates
are just vector subspaces of different dimensions ($d_{k i_k} = 1$, $d_{k'
i_{k'}} \gg 1$). For example, a macrostate may be defined as the set of all
microstates in which the center of mass of the apparatus pointer lies within
some given range of coordinates. The variables used to define the macrostate are
called \emph{collective} variables. For a macroscopic subsystem, the collective
variables are often described by commuting operators that \emph{approximate} the
values of noncommuting observables, such as the center-of-mass position and
momentum \cite{vonNeumann1955, VanKampen1954, dEspagnat1976, Omnes1992,
Omnes1994, Omnes1999a}.

Other important collective variables include the order parameters for states of
broken symmetry, which describe the emergence of distinct phases of matter
\cite{Anderson1984}. These order parameters exhibit a generalized ``rigidity''
property \cite{Anderson1984, Anderson1986} that is said to arise from a
``quantum protectorate'' \cite{LaughlinPines2000}. Such generalized rigidities
are useful in defining the ``pointer'' variables of actual or hypothetical
measuring apparatuses \cite{Anderson1984, Anderson1986}.

The different microstates within a given macrostate can be labeled using a set
of \emph{internal} variables. For example, suppose a hydrogen atom is known to
occupy a subsystem consisting of a given volume in space. Suitable collective
and internal coordinates are then the atomic center-of-mass position and the
relative electron--proton separation, respectively. In textbooks, the Hilbert
space of the hydrogen atom is usually decomposed into a tensor product of
independent center-of-mass and relative coordinate spaces. This cannot be done
exactly in the present example, because in a finite-volume subsystem, the set of
possible relative coordinates depends on the position of the center of mass.
Decomposing a subsystem into a tensor product of collective and internal
subsystems therefore represents an approximation that may be emergent in certain
limiting cases (such as, in this example, bound states of the hydrogen atom that
are not too close to the subsystem boundaries).

If such a tensor-product decomposition is emergent, the internal subsystem is
sometimes viewed as an \emph{internal environment} of the collective subsystem.
The collective subsystem then has both external and internal environments. The
notion of an internal environment can also be used in a loose sense even when no
such tensor-product decomposition is performed.

\subsection{Integrated and isolated subsystems}

\label{sec:integrated_isolated}

A subsystem is said to be \emph{isolated} if it interacts very weakly with other
subsystems, so that its time evolution is nearly unitary, apart perhaps from
occasional strong interactions with a measuring apparatus. A \emph{closed}
system, by contrast, interacts with nothing else at all.

Most macroscopic subsystems are practically impossible to isolate
\cite{Zeh1970}; isolation is therefore typically associated with microscopic
subsystems. Exceptions to this rule include the subsystems associated with the
phase order parameter of a superfluid or superconductor. A subsystem that is
never isolated on any relevant timescale is referred to here as an
\emph{integrated} subsystem. Integrated subsystems are important for
irreversibility in statistical mechanics \cite{Blatt1959} and the quantum theory
of measurement \cite{Zeh1970, Peres1986, Peres1988b}. Sometimes it is convenient
to use the words ``macroscopic'' and ``microscopic'' as imprecise substitutes
for ``integrated'' and ``isolated,'' but the latter labels are used
preferentially in the discussion that follows.

\subsection{Ideal observers}

\label{sec:ideal_observer}

One motivation sometimes given for the definition of macrostates is that a
protoscientist or ``macroscopic observer'' \cite{vonNeumann1955secV4} is assumed
to be capable of distinguishing pure states in the subspace defined by $P_i$
from those in a different subspace $P_j$, but incapable of distinguishing
different states within a given subspace $P_i$. The set (\ref{eq:P_set}) could
thus be regarded as characterizing the capabilities of such an observer
\cite{vonNeumann1955secV4}. However, if the theory is to achieve the goal of
being weakly objective (cf.\ Sec.\ \ref{sec:science_contradictory}) rather than
subjective, the macrostate projectors $P_i$ should be far more fine-grained than
the resolving capabilities of any protoscientist using the theory
\cite{GellMannHartle1990}.

The projectors in a weakly objective theory can thus be taken to
characterize the capabilities of an \emph{ideal} observer whose powers of
perception far exceed those of any existing protoscientist. Such an ideal
observer is referred to here as a Gell-Mann--Hartle \footnote{A GMH demon is,
roughly, an observer capable of accessing the information contained in histories
belonging to a set that approaches what Gell-Mann and Hartle have called a
``maximal'' \cite{GellMannHartle1990, Hartle1991a} or ``full''
\cite{GellMannHartle1991, GellMannHartle1993} set.} (GMH) demon. This demon is
of course imaginary. But the GMH demon, unlike Maxwell's or Laplace's demon,
could exist, at least in principle, within the confines of ordinary quantum
mechanics.

In this paper, the word ``information'' generally refers to the information
accessible to a GMH demon (thus answering Bell's question \cite{Bell1990}
``\emph{whose} information?''\ \footnote{The answer to Bell's other question,
``information about \emph{what}?'', is that the information is about the
emergent outcomes discussed in Secs.\ \ref{sec:emergence_outcomes} and
\ref{sec:partial_reduction}.}). Of course, the only actual information is that
held by individual protoscientists. The GMH demon is just an idealization useful
for describing the collective information potentially accessible to the
community of protoscientists in a manner consistent with the goal of weak
objectivity. The relationship between the GMH demon and the protoscientists is
discussed further in Sec.\ \ref{sec:weak_objectivity}.

\subsection{Restriction to integrated subsystems}

\label{sec:integrated_subsystems}

The definition (\ref{eq:compound_projector}) of $P_i$ as a product of subsystem
projectors implicitly assumes that the reduction process consists of $n$
reductions applied simultaneously to each of the individual subsystems. However,
as discussed below in Sec.\ \ref{sec:emergent_noninterference}, the time
interval $\Delta t$ between reductions in a closed system must satisfy
\begin{equation}
\Delta t \gg \tau_{\text{dec}} , \label{eq:decoherence_time}
\end{equation}
in which 
$\tau_{\text{dec}}$ is the decoherence time. This gives rise to a potential
problem, in that different subsystems $k$ generally have different decoherence
times $\tau_{\text{dec}} (k)$. The timescale $\tau_{\text{dec}}$ in Eq.\
(\ref{eq:decoherence_time}) for the simultaneous subsystem reduction generated
by $P_i$ must therefore be understood as the upper bound
\begin{equation}
\tau_{\text{dec}} = \max_{k} \tau_{\text{dec}} (k) . \label{eq:decoherence_max}
\end{equation}
Although $\tau_{\text{dec}} (k)$ is very small for typical macroscopic
subsystems \cite{Schlosshauer2007, JoosZeh2003, Zurek1991, Zurek2003,
HarocheRaimond2006}, it is very large (effectively infinite) for
isolated subsystems. For this reason, the compound projector
(\ref{eq:compound_projector}) is initially assumed to include only integrated
subsystems, and $P_{k i_k}$ is assumed to refer only to macrostates. The special
case of subsystem decompositions involving isolated subsystems will be
treated separately in Sec.\ \ref{sec:partial_reduction}.

Note that this temporary restriction to integrated subsystems does not mean
that Schr\"odinger-cat-type problems cannot be described; it only means that any
isolated subsystem must be absorbed into the definition of some integrated
subsystem, rather than being treated as a separate entity.

\subsection{Non-projective measurements}

\label{sec:non_projective_measurements}

Measurements can also be described by operators that are not projectors, as part
of the quantum operations formalism for open systems \cite{Peres1995,
NielsenChuang2000}. However, such measurement operators are always equivalent to
projectors acting in a larger Hilbert space. Since the present work deals only
with closed systems, all measurements are taken here to be projective for
simplicity. Nevertheless, the effect of such a measurement on an isolated
subsystem of the closed system generally cannot be described using projection
operators; this topic will be discussed further in Sec.\
\ref{sec:isolated_subsystems}.

\subsection{Is reduction ``real''?}

In the modern language of the histories formalism \cite{Griffiths1984,
Griffiths2002, Omnes1992, Omnes1994, Omnes1999a, GellMannHartle1990,
GellMannHartle1993, GellMannHartle2014, RiedelZurekZwolak2016}, the L\"uders
reduction is considered to be not a ``real'' process but
just a convenient way of describing the probabilities that are relevant to a
protoscientist within the given closed system \cite{Hartle1991a, Hartle1993b}.
This way of thinking can be traced back to Everett \cite{Everett1957,
Everett1973, Saunders2010} and arguably even to von Neumann himself
\cite{Becker2004}. However, any such ``reduction-free'' description must still
explain how it is possible for a protoscientist to perceive that its experiences
are described by the mathematics of reduction. Hence, the question of whether
reduction should be called ``real'' is irrelevant; the key question is whether
reduction is emergent.

\section{Emergence of noninterference}

\label{sec:emergent_noninterference}

The first step toward answering the latter question is to tackle the
interference problem. This problem arises in the quantum-mechanical description
of histories, the simplest example of which is the double-slit experiment. By
convention, a ``history'' in quantum mechanics is often formally defined as a
time-ordered product of Heisenberg-picture projection operators
\cite{Griffiths1984, Griffiths2002, Omnes1992, Omnes1994, Omnes1999a,
GellMannHartle1990, GellMannHartle1993, GellMannHartle2014,
RiedelZurekZwolak2016}. Suppose, however, we define a history simply as a
sequence of L\"uders reductions applied at times $t_1 < t_2 \cdots < t_f$ to
some initial state $\rho (t_0)$, where $t_0 \le t_1$ and $\rho (t)$ evolves
unitarily between reductions. We can then construct a probability distribution
for these histories by applying the rules of classical probability theory to the
conditional probabilities (\ref{eq:weight}), without any need for the apparatus
of decoherence functionals \cite{GellMannHartle1990, GellMannHartle1993}. There
is also no need for consistency conditions \cite{Griffiths1984, Griffiths2002,
Omnes1992, Omnes1994, Omnes1999a}, because the sum rules of classical
probability theory are satisfied automatically by construction.

However, this sequence of reductions can be used to describe a \emph{closed}
system only if the reductions do not interfere with each other. This means that,
if we calculate the probability of a given outcome at the final time $t_f$,
conditioned on the given initial state $\rho (t_0)$ but on no other information,
we should get the \emph{same} probability as if the reductions at the
intermediate times $t_1, t_2, \ldots, t_{f-1}$ were not performed at all. Such a
complete lack of interference is never \emph{exactly} true for any problem of
practical relevance, but it becomes a very good approximation in the limit
(\ref{eq:decoherence_time}), in which $\Delta t = \Delta t_j = t_j - t_{j-1}$ is
the time interval between reductions. The decoherence time $\tau_{\text{dec}}$
is state-dependent, but it is very fast for typical macroscopic subsystems. The
resulting \emph{emergence of noninterference} on the timescale
(\ref{eq:decoherence_time}) is the fundamental practical lesson of decoherence
theory \cite{Schlosshauer2007, JoosZeh2003, Zurek1991, Zurek2003,
HarocheRaimond2006}. Within the limits defined by Eq.\
(\ref{eq:decoherence_time}), reduction processes at intermediate times can be
inserted or deleted without having any significant effect on the overall
probability distribution.

Of course, the emergence of noninterference depends very sensitively on the
choice of projector set $\mathcal{P} (t_j)$. The problem of defining suitable
criteria for the selection of this set is known as the preferred-basis problem.
Methods for addressing this problem are discussed in Sec.\
\ref{sec:preferred_basis}.

As a summary of what has just been done, the rules for combining probabilities
in classical probability theory were injected (from out of nowhere) into the
formalism as a possible candidate for emergence. An appeal to the results of
decoherence theory was then used to establish that such emergence (i.e., of the
rules for combining probabilities) does indeed occur in the limit
(\ref{eq:decoherence_time}), subject to the solution of the preferred-basis
problem.

However, this emergence of noninterference does not imply the emergence of all
of classical probability theory, because it does not imply the emergence of
outcomes. The reason for this that noninterference is a property of probability
distributions for histories, whereas the outcome of the reduction process is a
quantum state. As shown below, the change of state during the L\"uders reduction
is generally not small enough to be described as emergent.

\section{Testing candidates for the emergence of outcomes}

\label{sec:testing_outcomes}

To test for the emergence of outcomes, we need to determine how much the state
changes upon reduction. A convenient measure for this is the trace distance
\cite{NielsenChuang2000, Bengtsson2017}
\begin{equation}
D(\rho, \sigma) = \frac12 \tr \abs{\rho - \sigma} , \label{eq:trace_distance}
\end{equation}
in which $\abs{X} = \sqrt{X^{\dagger} X}$.  
Some bounds on the relevant distances for the L\"uders reduction are evaluated
in Appendix \ref{app:reduction_pure} for the case in which $\rho$ is a pure
state (i.e., $\rho^2 = \rho$). As shown there, neither $D (\rho, \hat{\rho})$
nor $D (\rho, \rho_i)$ is small for the general case in which more than one
value of $w_i$ is significant. This was to be expected, since otherwise the
measurement problem would be trivial.

\subsection{Emergence of determinism}

\label{sec:emergent_determinism}

Consider, however, the special case in which one outcome, say $i = 1$, is
dominant: $w_1 = 1 - \epsilon$, $\epsilon \ll 1$. As shown in Appendix
\ref{app:reduction_pure}, $D(\rho, \hat{\rho})$ and $D(\rho, \rho_1)$ both
vanish in the limit $\epsilon \to 0$, whereas $D(\rho, \rho_{i \ne 1})$
approaches 1. Thus, in this limit, the states $\rho$ and $\rho_1$ are the same.
If we agree to ignore all outcomes except $i = 1$, we can say that deterministic
behavior has emerged. In this limit $w_i$ is either 0 or 1, so this corresponds
to the emergence of Boolean logic rather than of probability theory
\cite{Omnes1992, Omnes1994, Omnes1999a}.

\subsection{Collective variables and quasiclassical dynamics}

\label{sec:collective}

The conditions needed to make this happen have been studied by Omn\`es
\cite{Omnes1992, Omnes1994, Omnes1999a}. It is possible only for certain special
states $\rho$ and projector sets $\mathcal{P}$, the latter of which can be
chosen so as to minimize $D (\rho, \hat{\rho})$. The solution of this variation
problem is beyond the scope of the present paper.

However, qualitatively correct solutions can often be obtained without going
into the details of the variation problem. The key is to choose a set of
collective variables that evolve slowly in time, so as to obtain approximately
classical behavior \cite{vonNeumann1955, VanKampen1954, Omnes1992, Omnes1994,
Omnes1999a, GellMannHartle1990, GellMannHartle1993, GellMannHartle2014}. In
general, the concepts of symmetry breaking and order parameters play a crucial
role in the selection of collective variables (cf.\ Sec.\
\ref{sec:microstates_macrostates}). However, in common ``hydrodynamic''
situations the appropriate variables might simply consist of the number of
particles in a small but macroscopic volume together with the center-of-mass
position, total momentum, and internal energy of those particles. Once the
quasiclassical variables have been chosen, subsystem projectors
(\ref{eq:subsystem_projector0}) can be defined corresponding to some given
ranges of these variables and combined to obtain the overall projector
(\ref{eq:compound_projector}).

The resulting emergence of quasiclassical determinism is valid for systems whose
behavior can be predicted accurately using classical mechanics \cite{Omnes1992,
Omnes1994, Omnes1999a}. It is, however, not valid for classically chaotic
systems or for ``Schr\"odinger cat'' situations in which, say, $w_1 = w_2 =
1/2$. In these cases, $D (\rho, \hat{\rho})$ and $D (\rho, \rho_i)$ are not
small and we must turn to other methods.

\subsection{Change of metric}

\label{sec:change_metric}

The crucial step in all cases of emergence is the choice of what to ignore.
Given that the physical predictions of quantum mechanics are all based on
probabilities, this choice can be made explicit by defining the quantity
\begin{equation}
D_{\mathcal{Q}} (\rho, \sigma) = \sup_{P \in \mathcal{Q}} \abs{\tr [P (\rho -
\sigma)]} , \label{eq:variation_distance}
\end{equation}
in which $\mathcal{Q}$ is some given set of projectors \footnote{$\mathcal{Q}$
can also be defined as a set of positive operators \cite{NielsenChuang2000}:
$\mathcal{Q} = \{ P : 0 \le P \le 1 \}$. However, this choice is not used
here.}. The physical meaning of this definition can be seen by noting that for
any two density operators $\rho$ and $\sigma$,
\begin{equation}
D_{\mathcal{Q}} (\rho, \sigma) = 0 \ \Leftrightarrow \ \forall P \in \mathcal{Q} , \
\tr (P \rho) = \tr (P \sigma) . \label{eq:D_Z_zero}
\end{equation}
The condition $D_{\mathcal{Q}} (\rho, \sigma) = 0$ is thus identical to the
statement that $\rho$ and $\sigma$ give rise to the same probabilities for
measurements of any projectors $P \in \mathcal{Q}$. If measurements are
restricted to observables that are linear combinations of these projectors, the
density operators $\rho$ and $\sigma$ would therefore be operationally
indistinguishable \cite{Weidlich1967, Fine1970}.

The definition (\ref{eq:variation_distance}) is related to the trace distance
(\ref{eq:trace_distance}) by \cite{NielsenChuang2000, Bengtsson2017}
\begin{equation}
D_{\mathcal{Q}} (\rho, \sigma) \le D (\rho, \sigma) ,
\label{eq:trace_distance_inequality}
\end{equation}
in which the equality holds for all $\rho$ and $\sigma$ if and only if
$\mathcal{Q}$ is dense in $\mathcal{Q}_0$, the set of all projectors. If this is
not true, $D_{\mathcal{Q}} (\rho, \sigma)$ is called a pseudometric, because it
satisfies all of the criteria for a metric except the requirement that
$D_{\mathcal{Q}} (\rho, \sigma) = 0$ implies $\rho = \sigma$. It can be
converted into a metric if we regard it as defining a distance between
equivalence classes, where $\rho$ and $\sigma$ belong to the same equivalence
class if $D_{\mathcal{Q}} (\rho, \sigma) = 0$. The subject of equivalence
classes is discussed below in Sec.\ \ref{sec:emergence_outcomes}.

\subsection{Macroscopically local observables}

\label{sec:macroscopically_local}

A common choice for $\mathcal{Q}$ is the set $\mathcal{L}$ of projectors
contained in the algebra of local (or quasilocal) ``observables.'' In algebraic
quantum mechanics, such observables are defined as operators with compact
support in coordinate space (or, for relativistic problems, Minkowski space)
\cite{HaagKastler1964, Haag1996}. This definition is appropriate for
mathematical investigations of limiting behavior \cite{HaagKastler1964,
Haag1996, Hepp1972, Prosperi1974, Bell1975}, but it is not suitable for use in
any realistic theory of measurement. For the latter purpose, it is better to
work with the algebra of \emph{macroscopically} local observables, defined as
those with a \emph{fixed} upper bound on the spatial extent of nonlocality
\cite{PeresRosen1964, Gottfried1966}.

Gottfried has given an insightful analysis of the measurement problem in which
this set is applied to the case of a Stern--Gerlach experiment that
measures the spin of an atom \cite{[] [{, Sec.\ 20.}] Gottfried1966}. In his
approach, the initial state $\rho (t_a)$ at time $t_a$ is the outcome of a
state-preparation or reduction process. This state evolves in accordance with
Eq.\ (\ref{eq:unitary}) until $t_b = t_a + \Delta t$, at which time $\rho (t_b)$
is reduced to obtain $\hat{\rho} (t_b)$. The purpose of Gottfried's analysis is
to show that this reduction is justified by the fact that
\begin{equation}
D_{\mathcal{L}} [\rho (t_b), \hat{\rho} (t_b)] \approx 0 
\label{eq:local_emergence}
\end{equation}
for sufficiently large $\Delta t$, in which $\mathcal{L}$ is the set of
projectors contained in the algebra of macroscopically local observables. In
order to achieve this, the projector set $\mathcal{P}_b = \mathcal{P} (t_b)$ is
chosen so as to partition the center of mass of the atom into macroscopically
distinct regions. This could be regarded as the solution of a variation problem,
in which $\mathcal{P}_b$ is varied so as to minimize $D_{\mathcal{L}} [\rho
(t_b), \hat{\rho} (t_b)]$.

If we accept Gottfried's claim that ``all known observables'' are
macroscopically local \cite{Gottfried1966}, the result
(\ref{eq:local_emergence}) would seem to be an important first step (but not a
complete solution---see Sec.\ \ref{sec:emergence_outcomes}) towards establishing
the emergence of outcomes, at least in this special case. However, this claim
made Bell ``quite uncomfortable'' \cite{Bell1990}, and for good reason. The
explicit reference to locality in coordinate space is a statement about what is
observable to \emph{humans}. The preferred-basis problem is thereby not solved,
it is merely eliminated by fiat.

However, as Anderson has stressed, the ``pointer'' of a measuring apparatus need
not point to anything in coordinate space; it could, at least in principle, be
based on a generalized concept of ``rigidity'' involving, say, the phase
variable of liquid helium \cite{Anderson1984, Anderson1986}. Such phases can
only be perceived by humans if they are measured by a more traditional apparatus
whose pointer is macroscopically local in coordinate space \cite{[{Peres points
out that ``a macroscopic superconducting current is \emph{not} acceptable as a
measuring instrument, unless it is coupled to some monitoring device with
numerous degrees of freedom.'' }] [] Peres1980d}, but Anderson speculates that
they could perhaps be perceived directly by a computer ``made entirely out of
Josephson junctions'' \cite{Anderson1986}.

The reference to macroscopic locality thus introduces a subjective element into
the theory. It is desirable to see whether this element of subjectivity can be
eliminated by defining $\mathcal{Q}$ in a different way.

\subsection{Are local ``observables'' actually observed?}

\label{sec:local_observables_unobservable}

Another reason to be uncomfortable with relying on the set $\mathcal{L}$ is that
the projectors in $\mathcal{L}$ do not always correspond directly to the outcome
of any experiment. The experimentally accessible quantities in quantum mechanics
are (with some qualifications to be explored in detail below) the reduced states
$\rho_i$ and their associated probabilities $w_i$. These quantities are directly
related to the projectors $P_i$ appearing in the set $\mathcal{P}_b$ referred to
above. However, these projectors do not necessarily belong to $\mathcal{L}$; as
noted above, the set $\mathcal{P}_b$ is defined by the requirement that it
minimize $D_{\mathcal{L}} [\rho (t_b), \hat{\rho} (t_b)]$, not by membership in
$\mathcal{L}$. In the language of algebraic quantum mechanics, the set
$\mathcal{P}_b$ that defines the basis in which $\rho (t_b)$ and $\hat{\rho}
(t_b)$ become equivalent in the sense of Eq.\ (\ref{eq:local_emergence}) belongs
to the set of \emph{macroscopic classical} observables, not to the set of
\emph{local} observables \cite{Hepp1972, Prosperi1974, Bell1975}. The former is
generally not a subset of the latter \cite{Hepp1972, Prosperi1974, Bell1975}.

Hence, the choice $\mathcal{Q} = \mathcal{L}$ represents somewhat of a diversion
from the physical motivation given for the introduction of $D_{\mathcal{Q}}
(\rho, \sigma)$, because the probabilities associated with projectors in
$\mathcal{L}$ are not necessarily probabilities of any experimental outcome. The
``observables'' in $\mathcal{L}$ are thus actually a class of (partially)
\emph{hidden variables} introduced in order to facilitate the solution of the
measurement problem. But this calls into question the significance of the
results obtained from minimization of $D_{\mathcal{L}} (\rho, \sigma)$.

It is therefore of interest to see whether the problem can be reformulated in
terms of a set $\mathcal{Q}$ that is more closely connected to experimental
outcomes. This can be achieved by replacing the criterion of macroscopic
locality with that of dynamical stability, as outlined below.

\subsection{Dynamical stability}

\label{sec:dynamical_stability}

The concept of dynamical stability has played an important role in decoherence
theory from the very beginning \cite{Zeh1970, Zeh1971b}. It has roots in the
criteria of stability \cite{[] [{; English translation in }] Schrodinger1926c,
*Schrodinger1982_p41_44} and predictability \cite{EPR1935} used to define
notions of ``reality'' by the founders of quantum mechanics. Dynamical stability
can be implemented here by replacing the choice $\mathcal{Q} = \mathcal{L}$ in
Sec.\ \ref{sec:macroscopically_local} with $\mathcal{Q} = \mathcal{P}_a$, where
$\mathcal{P}_a = \mathcal{P} (t_a)$ is the set of projectors used in the
reduction process that generated the initial state $\rho (t_a)$.

The reason why this corresponds to a criterion of dynamical stability can be
seen by noting that $D_{\mathcal{P}_a} [\rho (t_b), \hat{\rho} (t_b)] = 0$ if
$\mathcal{P}_b = \mathcal{P}_a$, due to the idempotence of the L\"uders
reduction process.  In general, $\mathcal{P}$ must depend on time,
so we have at best
\begin{equation}
D_{\mathcal{P}_a} [\rho (t_b), \hat{\rho} (t_b)] \approx 0 
\label{eq:dynamical_stability}
\end{equation}
when $\mathcal{P}_b \approx \mathcal{P}_a$. This corresponds to the case of
projectors that vary slowly with time, in the sense that the interval $\Delta t$
between reductions can be chosen to satisfy [cf.\ Eq.\
(\ref{eq:decoherence_time})]
\begin{equation}
\tau_{\text{dec}} \ll \Delta t \ll \tau_{\mathcal{P}} , \label{eq:slow_time}
\end{equation}
where $\tau_{\mathcal{P}}$ is the timescale for changes in $\mathcal{P}$. Hence,
the criterion (\ref{eq:dynamical_stability}) can be used to select projector
sets $\mathcal{P}$ that are stable over time intervals long in comparison to the
decoherence time.

Unlike the situation in Sec.\ \ref{sec:macroscopically_local}, here the set
$\mathcal{P}_b$ is not regarded as a quantity to be varied in order to minimize
$D_{\mathcal{P}_a} [\rho (t_b), \hat{\rho} (t_b)]$, because that would yield
only the trivial time-independent solution $\mathcal{P}_b = \mathcal{P}_a$. The
selection of $\mathcal{P}_b$ (i.e., the solution of the preferred-basis problem)
is instead governed by a separate variational principle.

For the moment, the preferred-basis problem is set aside, to be taken up
again in Sec.\ \ref{sec:preferred_basis}. The question of immediate concern is,
rather, what precisely is the outcome of the reduction process?

\section{Emergence of outcomes in integrated subsystems}

\label{sec:emergence_outcomes}

\subsection{Density operators have too much information}

\label{sec:incomplete_information}

This question would seem to have an obvious answer: The outcome at time $t_j$ is
the reduced state $\rho_i$ in Eq.\ (\ref{eq:rho_i_Lu}). The problem with this
answer is that we never actually know what $\rho_i (t_j)$ is; we only know that
it is some state that belongs to the manifold
\begin{equation}
\mathcal{M}_i (t_j) = \{ \sigma \mid P_i (t_j) \sigma = \sigma P_i (t_j) =
\sigma \} .
\end{equation}
Whatever this state is, it then evolves unitarily until the time of the next
reduction, yielding another unknown state $\rho (t_{j+1})$. The only connection
between the pre-reduction state $\rho (t_{j+1})$ and experiment is the set $\{
w_i (t_{j+1}) \}$, in which $w_i (t_{j+1})$ is the predicted probability
(\ref{eq:weight}) of obtaining an outcome in $\mathcal{M}_i (t_{j+1})$.

The outcome of an experiment is therefore ill-defined if we take the quantum
state to be a density operator $\rho$. The crux of the problem is that $\rho$
contains too much information. As noted in Sec.\ \ref{sec:change_metric}, the
operator $\rho$ would be completely determined if we knew the value of $\tr (P
\rho)$ for a complete set of projectors $P$ \cite{dEspagnat1976, [] [{, p.\
49.}] GottfriedYan2003}. This information can be obtained for an assembly of
open isolated systems by the technique known as quantum-state tomography, as
described in Sec.\ \ref{sec:quantum_state_tomography}.

However, we \footnote{Here the word ``we'' includes all protoscientists and the
GMH demon.} never have access to complete information about the state of a
closed system \cite{Everett1973p98}. In such a system, which is by definition
unique, we can only obtain information about $\rho$ at the reduction times
$t_j$. The information thus obtained is limited to the outcomes $\mathcal{M}_i$,
which can be used to test the probabilities $w_i$ calculated from Eq.\
(\ref{eq:weight}). That is, our experimental information can only make contact
with the values of $\tr (P \rho)$ for all $P \in \mathcal{P}$, in which
$\mathcal{P}$ is the set (\ref{eq:P_set}) used to perform the reduction. This
set is not fixed \emph{a priori}; it is contingent upon the dynamical solution
of the preferred-basis problem (see Sec.\ \ref{sec:preferred_basis}). But at any
given time, we have access to the values of $\tr (P \rho)$ only for one
particular set $\mathcal{P}$, not for all conceivable sets $\mathcal{P}$. (This
is also true for an ensemble of open integrated systems, as discussed in Sec.\
\ref{sec:emergent_statistics}.) This is the basic limitation imposed by Bohr's
principle of complementarity \cite{Bohr1935}, which can be summarized in the
statement that ``unperformed experiments have no results'' \cite{Peres1978}.

\subsection{Equivalence classes}

\label{sec:equivalence_classes}

If all of the information we have about $\rho$ at some given time is contained
in the set of probabilities $\{ w_i \}$ in Eq.\ (\ref{eq:weight}), then we
cannot distinguish $\rho$ from any other state $\sigma$ that gives rise to the
same set $\{ w_i \}$. This indistinguishability can be expressed by defining an
equivalence relation between density operators. We say that $\rho$ and $\sigma$
are equivalent with respect to the set $\mathcal{P}$, denoted $\rho \sim \sigma
\, (\mathcal{P})$, if $D_{\mathcal{P}} (\rho, \sigma) = 0$ or [cf.\ Eq.\
(\ref{eq:D_Z_zero})]
\begin{subequations}
\label{eq:equivalence_compound}
\begin{equation}
\tr (P \rho) = \tr (P \sigma) \ \forall P \in \mathcal{P} .
\label{eq:equivalence_relation_compound}
\end{equation}
The equivalence class of $\rho$ with respect to $\mathcal{P}$ is then
defined as the set
\begin{equation}
\eqcl{\rho} = \{ \sigma \mid \sigma \sim \rho \, (\mathcal{P}) \} .
\label{eq:equivalence_class_compound}
\end{equation}
\end{subequations}
The motivation for this is that an ideal but physically possible observer---a
GMH demon---is assumed to be capable of distinguishing states in the manifold
$\mathcal{M}_i$ from those in another manifold $\mathcal{M}_j$, but incapable of
distinguishing different states within $\mathcal{M}_i$. The set (\ref{eq:P_set})
can thus be regarded as characterizing the capabilities of the GMH demon. All
states $\sigma \in \eqcl{\rho}$ are indistinguishable to the demon. If we wish
to obtain a description that is logically consistent with the capabilities of
even such an ideal observer, we should therefore redefine the quantum state as
the equivalence class $\eqcl{\rho}$ itself \cite{vonNeumann1955secV4, Jauch1964,
Jauch1968}.

Note that the definition of quantum states as equivalence classes is already
well established as a basic feature of standard quantum mechanics. For example,
in ordinary wave mechanics, the so-called ``wave function'' is actually an
equivalence class of functions that are equal \emph{almost} everywhere (i.e.,
except on sets of Lebesgue measure zero) \cite{[] [{, p.\ 42.}] Isham1995}. More
generally, the quantum state is defined as a probability measure on a set of
projectors \cite{Gleason1957}. Under certain conditions, this definition leads
to a one-to-one correspondence between quantum states and density operators
\cite{Gleason1957}. However, in systems with exact superselection rules, this
correspondence is broken, and the quantum state can only be identified as an
equivalence class of density operators \cite{[] [{, pp.\ 49 and 57.}]
Beltrametti1981}. It is therefore not surprising that the same should be true
in systems with emergent environment-induced superselection rules
\cite{Zurek1982, Zurek1991}.

The equivalence class (\ref{eq:equivalence_compound}) has several noteworthy
properties. For any density operator $\rho$ we have
\begin{equation}
\eqcl{\hat{\rho}} = \eqcl{\rho} ,  \label{eq:rho_hat_eqcl}
\end{equation}
in which $\hat{\rho}$ is the L\"uders reduced state defined in Eqs.\
(\ref{eq:reduction_a}) and (\ref{eq:rho_i_Lu}). That is, $\rho$ and $\hat{\rho}$
are always equivalent. However, there are significant advantages to redefining
$\hat{\rho}$ at this point as \cite{vonNeumann1955secV4}
\begin{equation}
\hat{\rho} = \sum_{i} \frac{\tr (\rho P_i)}{\tr P_i} P_i , \label{eq:rho_hat}
\end{equation}
thereby extending von Neumann's original definition [see Eqs.\
(\ref{eq:reduction_a}) and (\ref{eq:rho_i_vN1})] to the case of $\tr P_i \ge 1$.
This definition still satisfies Eq.\ (\ref{eq:rho_hat_eqcl}), but it also
satisfies the much stronger condition
\begin{equation}
\eqcl{\rho} = \eqcl{\sigma} \ \Leftrightarrow \ \hat{\rho} = \hat{\sigma} ,
\label{eq:eqcl_representation}
\end{equation}
as shown in Appendix \ref{app:eqcl_representation}. Hence, under the new
definition, there is a one-to-one relationship between equivalence classes
$\eqcl{\rho}$ and reduced states $\hat{\rho}$. The von Neumann reduced state
(\ref{eq:rho_hat}) can therefore be used to represent the equivalence class
$\eqcl{\rho}$, subject to some qualifications about dynamics to be discussed
below in Sec.\ \ref{sec:modified_reduction_process}.

The uniqueness of the reduced state (\ref{eq:rho_hat}) is also reflected in the
maximum-entropy property
\begin{equation}
S(\hat{\rho}) \ge S (\sigma) \ \forall \sigma \in \eqcl{\rho} ,
\label{eq:maximum_entropy}
\end{equation}
in which
\begin{equation}
S (\rho) = - \tr (\rho \ln \rho)  \label{eq:entropy_vN}
\end{equation}
is the von Neumann entropy \cite{vonNeumann1955}. The property
(\ref{eq:maximum_entropy}) is a consequence of Eq.\
(\ref{eq:eqcl_representation}) and the inequality \cite{vonNeumann1955secV4}
\begin{equation}
S(\hat{\rho}) \ge S (\rho) , \label{eq:increased_entropy}
\end{equation}
the latter of which also holds for the L\"uders reduced state
\cite{NielsenChuang2000}. The equality in (\ref{eq:maximum_entropy}) or
(\ref{eq:increased_entropy}) holds if and only if $\sigma = \hat{\rho}$ or $\rho
= \hat{\rho}$ \cite{vonNeumann1955secV4}, respectively.

Further properties of the definition (\ref{eq:rho_hat}) are easily derived.
For the special case of a projector in the set $\mathcal{P}$, we get simply
\begin{equation}
\hat{P}_i = P_i , \label{eq:P_i_hat}
\end{equation}
which implies that the operation (\ref{eq:rho_hat}) is idempotent:
\begin{equation}
\hat{\hat{\rho}} = \hat{\rho} . \label{eq:reduction_idempotent}
\end{equation}
Also, for any pure state $\chi_i$ that belongs to the manifold $\mathcal{M}_i$
defined by $P_i$, we have
\begin{equation}
\hat{\chi}_i = \frac{P_i}{\tr P_i} . \label{eq:pure_reduced}
\end{equation}
Given that the equivalence class $\eqcl{\rho}$ can be represented mathematically
by the reduced state $\hat{\rho}$, these results indicate that the basic
dynamical entity of the theory is not the pure state but the projector $P_i$.
This projector represents its own equivalence class, due to Eq.\
(\ref{eq:P_i_hat}).

\subsection{Modified reduction process}

\label{sec:modified_reduction_process}

If we follow this approach, the L\"uders reduced state (\ref{eq:rho_i_Lu}) in
the reduction process (\ref{eq:reduction}) should be replaced by
\begin{align}
\rho_i = \frac{P_i}{d_i} , \label{eq:rho_i_vN2}
\end{align}
in which $d_i = \tr P_i$. This is just a generalization of von Neumann's reduced
state (\ref{eq:rho_i_vN1}) to the case $d_i \ge 1$. The reduction process
defined by Eqs.\ (\ref{eq:reduction}) and (\ref{eq:rho_i_vN2}) could be regarded
as implicit in von Neumann's work, because he devoted a significant portion of
Ref.\ \cite{vonNeumann1955} to a study of the implications of the reduced state
(\ref{eq:rho_hat}) for the increase of entropy predicted by the second law of
thermodynamics \cite{vonNeumann1955secV4}. However, von Neumann always wrote his
reduction process in the form of Eqs.\ (\ref{eq:reduction}) and
(\ref{eq:rho_i_vN1}), with $d_i = 1$. To the best of my knowledge, the first
authors to write down Eqs.\ (\ref{eq:reduction}) and (\ref{eq:rho_i_vN2})
explicitly were Daneri, Loinger, and Prosperi \cite{Daneri1962} (see also Refs.\
\cite{Prosperi1971, Prosperi1974}). In view of this ambiguous heritage, I shall
refer to this process as the vN--DLP reduction.

There is a subtlety here in that different members of an equivalence class are
\emph{not} equivalent with respect to dynamics, because the projectors $P_i$
generally do not commute with the Hamiltonian $H$ \cite{vonNeumann1955secV4}.
Different members of an initial equivalence class therefore evolve differently
during the time interval between reductions, and at the time of the next
reduction they will generally belong to different equivalence classes. This
\emph{dynamical} lack of information can be accounted for by introducing a
probability distribution over the different members $\sigma_i \in \eqcl{\rho_i}$
of the equivalence class $\eqcl{\rho_i}$ that defines the actual state of the
system at the time of reduction. However, the introduction of such a probability
distribution is equivalent to choosing to represent the reduced state
$\eqcl{\rho_i}$ by some other specific mixed state $\sigma_{i} \in
\eqcl{\rho_i}$ with $\sigma_{i} \ne \rho_i$.

Defining this probability distribution requires the introduction of some new
principle that is foreign to quantum mechanics \emph{per se}. One possible
candidate is the principle of maximum entropy, which in an information-theory
context represents the policy of being ``maximally noncommittal with regard to
missing information'' \cite{Jaynes1957a, *Jaynes1957b}. But according to Eq.\
(\ref{eq:maximum_entropy}) this corresponds to the choice $\sigma_i =
\hat{\rho}_i = \rho_i$, thus leading back once again to the formulation given in
Eq.\ (\ref{eq:rho_i_vN2}).  This choice is adopted in what follows.

\subsection{Emergence of classical probability}

The crucial question now is whether the reduction process defined by Eqs.\
(\ref{eq:reduction}) and (\ref{eq:rho_i_vN2}) is emergent. To address this
question, note that the essential information content of the equivalence class
(\ref{eq:equivalence_compound}) is contained in the set of probabilities $\{ w_i
\}$, which can be represented by a vector
\begin{equation}
\vect{w} = (w_1, w_2, \ldots) . 
\end{equation}
All states $\rho$ in the same equivalence class have the same $\vect{w}$ vector,
so $\vect{w}$ is a characteristic of $\eqcl{\rho}$ itself. The distance $\Delta
(\eqcl{\rho}, \eqcl{\rho'})$ between equivalence classes $\eqcl{\rho}$ and
$\eqcl{\rho'}$ can then be defined as the Kolmogorov distance
\cite{FuchsGraaf1999} or classical trace distance \cite{NielsenChuang2000}
between $\vect{w}$ and $\vect{w}'$:
\begin{subequations}
\label{eq:Delta_eqcl}
\begin{align}
\Delta (\eqcl{\rho}, \eqcl{\rho'}) & = \frac12 \norm{\vect{w} - \vect{w}'}_1 \\
& = \frac12 \sum_{i} \abs{w_{i} - w_{i}'} .
\end{align}
\end{subequations}
As noted above, $\eqcl{\rho}$ can also be represented mathematically by
$\hat{\rho}$. But $\hat{\rho}$ and $\hat{\rho}'$ commute (for fixed
$\mathcal{P}$), so the trace distance (\ref{eq:trace_distance}) between them is
identical to Eq.\ (\ref{eq:Delta_eqcl}):
\begin{equation}
D (\hat{\rho}, \hat{\rho}') = \frac12 \sum_{i} \abs{w_{i} - w_{i}'} .
\label{eq:distance_rho_hat}
\end{equation}
Hence, these two measures of distance are the same. The essential difference
between the representations $\hat{\rho}$ and $\vect{w}$ of $\eqcl{\rho}$ is that
$\hat{\rho}$ can be used as an initial condition for the unitary dynamics
(\ref{eq:unitary}), whereas $\vect{w}$ cannot.

From these results we see immediately that
\begin{equation}
\Delta (\eqcl{\rho}, \eqcl{\hat{\rho}}) = 0 ,  \label{eq:distance_zero}
\end{equation}
because $\eqcl{\rho} = \eqcl{\hat{\rho}}$. Thus, the redefinition of quantum
states as equivalence classes has led to the vanishing of the difference between
the entangled state (\ref{eq:unitary}) and the corresponding reduced state
(\ref{eq:rho_hat}). It should be noted that Eq.\ (\ref{eq:distance_zero}) has
not appeared here entirely out of the blue; it was preceded by the very similar
dynamical-stability condition shown in Eq.\ (\ref{eq:dynamical_stability}).
Hence, in defining equivalence classes such that the distance $\Delta
(\eqcl{\rho}, \eqcl{\hat{\rho}})$ vanishes exactly, we are only choosing to
neglect a small quantity that was already shown to be negligible according to
the criterion of dynamical stability.

The first stage (\ref{eq:reduction_a}) of the reduction process is therefore
emergent. But this is sufficient to establish the emergence of outcomes, because
the interpretation of $\hat{\rho}$ as a statistical mixture of distinct outcomes
can now be imposed on Eq.\ (\ref{eq:rho_hat}) without generating any noticeable
conflict with unitary dynamics---in which ``noticeable'' is defined with respect
to the metrics (\ref{eq:Delta_eqcl}) and (\ref{eq:distance_rho_hat}). It should
be stressed that this interpretation is \emph{imposed on} Eq.\
(\ref{eq:rho_hat}), not \emph{derived from} it. Bell has argued that it would be
more natural to interpret equations of the type (\ref{eq:rho_hat}) as a
simultaneous coexistence of the projectors $P_i$---i.e., to interpret the
summation in Eq.\ (\ref{eq:rho_hat}) as a classical ``and'' rather than a
classical ``or'' \cite{Bell1990}. But this argument only carries weight if we
assume that our goal is to \emph{derive} the emergent theory from the base-level
theory. It is not. This is never the goal in any true case of emergence, which
always requires reconciliation of contradictions between the two levels of
theory.

Here it may also be objected \cite{Bell1990} that this is ``not the
proof of a theorem, but a \emph{change of the theory}---at a strategically well
chosen point.'' In other words, we have only managed to win the game by moving
the goalposts. This is true, but it is also true that this is \emph{always} how
the game of emergence is played and won. Nothing ever emerges until we have
chosen what to ignore.

The interpretation of $\hat{\rho}$ as a statistical mixture means that the
Hilbert-space weights $w_i$ in Eq.\ (\ref{eq:weight}) have now emerged as
classical probabilities. However, because the closed system under consideration
is not a member of any ensemble, they are probabilities only in the Bayesian
sense of plausible reasoning in the face of incomplete information
\cite{Jaynes1989, Jaynes2003}, not in the ``frequentist'' sense of frequencies
in an ensemble. But in order to connect $w_i$ with experiment, it is
necessary to introduce ensembles and statistical frequencies. This connection
will be developed below in Sec.\ \ref{sec:emergent_statistics}.

The second stage (\ref{eq:reduction_b}) of the vN--DLP reduction accounts for
the selection of one outcome $\rho_i$ from among the many possibilities present
in $\hat{\rho}$. The distance (\ref{eq:Delta_eqcl}) corresponding to this
process is
\begin{equation}
\Delta (\eqcl{\hat{\rho}}, \eqcl{\rho_i}) = 1 - w_i .
\label{eq:reduction_distance_1}
\end{equation}
This becomes small only in the deterministic limit $w_i \approx 1$ considered
previously in Sec.\ \ref{sec:emergent_determinism}. It is not small for the
general case in which no value of $w_i$ is close to 1. But this simply means
that the change in a probability distribution is always significant whenever we
update it to account for the acquisition of new information (where \emph{new}
information is, by definition, considered to be unexpected). It has no bearing
on the emergence of outcomes, because the outcomes have already emerged.

\subsection{Meaning of the results}

\label{sec:meaning_of_results}

What is the meaning of the vN--DLP reduction process? Here an analogy with
statistical mechanics is again helpful. In the context of Gibbs's famous
ink-in-water example \cite{Gibbs1902}, the Schr\"odinger dynamics
(\ref{eq:unitary}) generates a mixing of these fluids. But due to the
conservation of the eigenvalues of $\rho$ by the unitary process
(\ref{eq:unitary})---analogous to the incompressibility of the fluids---it is
well known that the second law of thermodynamics can only be recovered if the
microscopic state of the fluid is repeatedly ``reduced'' by coarse graining or
some other type of information loss \cite{Jaynes1965, Peres1988b, Zeh2007}. The
L\"uders reduction is one example of this type of coarse graining.

However, if the state of the system is to be described by equivalence classes,
we should take the reduced states $\hat{\rho}$ and $\rho_i$ to be given not by
the specific expressions (\ref{eq:reduction_a}) and (\ref{eq:rho_i_Lu}) but by
the corresponding equivalence classes. The introduction of equivalence classes
thus represents a further stage of coarse graining, in which the excess
information contained in Eq.\ (\ref{eq:rho_i_Lu}) is discarded. In practice,
however, this cumbersome two-step coarse-graining process is jettisoned in favor
of the single definition (\ref{eq:rho_i_vN2}). It was precisely this question of
agreement with the second law of thermodynamics that motivated von Neumann to
consider the reduced state (\ref{eq:rho_hat}).

In the L\"uders reduction, some coherence between the components of the initial
quantum state $\rho$ generally survives the reduction process, as long as $\tr
P_i > 1$. For the quasiclassical projectors that describe our everyday
experiences, typically $\rho$ ``is not reduced very much'' \cite{Hartle1991a}.
However, in the vN--DLP reduction, the initial state $\rho$ is replaced in its
entirety by a normalized projector (\ref{eq:rho_i_vN2}) at each reduction. The
state prior to the last reduction is completely forgotten. In other words, all
information about what is going on inside the equivalence classes is renounced
at each step of the process. This is in better accord with the
information-theory principle of being ``maximally noncommittal with regard to
missing information'' \cite{Jaynes1957a, *Jaynes1957b}.

Of course, given that the projectors $P_i$ describe the capabilities of the GMH
demon, any protoscientist using the theory will describe the experimental
outcome as a \emph{partial} reduction that is intermediate between
(\ref{eq:reduction_a}) and (\ref{eq:reduction_b}). That is, the
empirical reduced state will be a statistical mixture of states $\rho_i$ that is
narrower than Eq.\ (\ref{eq:reduction_a}) but still includes many different
$\rho_i$.

Zeh has cautioned against interpreting the reduction process as a ``mere
increase of information'' \cite{JoosZeh2003, Zeh2007, Zeh2006, Zeh2011}, in the
sense of merely selecting an outcome from some preexisting ensemble. One reason
for this is that applying reduction to a closed system when decoherence is not
yet complete (i.e., when the ``measurement'' is still in practice reversible)
leads to results in conflict with experiment \cite{Furry1936a}. Another reason
is that the outcomes are not preexisting; they emerge dynamically from the
solutions of the preferred-basis problem (see Sec.\ \ref{sec:preferred_basis}). 

Bub has suggested that arguments for the equivalence of $\rho$ and $\hat{\rho}$
cannot solve the measurement problem, because $\rho$ and $\hat{\rho}$ are
equivalent ``only from the point of view of a specific observer''
\cite{Bub1968a}. However, in the present theory, this observer is the GMH demon,
thereby establishing the weak objectivity of the theory. The equivalence of
$\rho$ and $\hat{\rho}$ is also bolstered by the criterion of dynamical
stability developed in Sec.\ \ref{sec:dynamical_stability}. If one insists on
nothing less than strong objectivity, then it is true that the measurement
problem is insoluble. But as argued in Sec. \ref{sec:science_contradictory}, the
requirement of strong objectivity is incompatible with the scientific method
itself.

Von Neumann showed that the reduction defined by Eqs.\ (\ref{eq:reduction_a})
and (\ref{eq:rho_i_vN2}) can be written mathematically as a product of two
maximally fine-grained reductions of type (\ref{eq:rho_i_vN1}), in which the
rank-one projectors within each equivalence class are chosen to form
complementary sets and the time interval $\Delta t$ between reductions is set to
zero \cite{vonNeumann1955secV4}. Such a product has been called a ``coherence
destroying'' measurement \cite{AharonovBergmannLebowitz1964}. However, it must
be stressed that the vN--DLP reduction cannot be interpreted physically as a
sequence of two conventional measurements, because $\Delta t$ cannot be zero for
measurements of noncommuting observables \cite{Griffiths1984} [recall $\Delta t
\gg \tau_{\mathrm{dec}}$ in Eq.\ (\ref{eq:slow_time})]. The limit $\Delta t \to
0$ is a singular limit that gives rise to new physics---in this case, a
conventional reduction followed by a renouncement of all information that would
distinguish between the states within an equivalence class.

It is interesting to note that when the reduction process is defined by Eqs.\
(\ref{eq:reduction}) and (\ref{eq:rho_i_vN2}), both the quantum state and all
observables of the type (\ref{eq:observable}) are functions of the projectors
$P_i$. The emergence of outcomes in a set of integrated subsystems thus leads to
a convergence between Schr\"odinger's concept of state and Heisenberg's concept
of observable, in a manner highly reminiscent of classical mechanics.

In a certain sense this may be considered as a trivialization of the measurement
problem \cite{[] [{, p.\ 338.}] Fuchs2011}, given the longstanding focus on the
question of how to deal with the freedom of the experimenter to measure any of
several possibly noncommuting observables \cite{Bohr1935, PeresZurek1982,
Peres1986c, Peres1995, Primas1990a0}. However, it should be kept in mind that the
projectors $P_i$ are not arbitrary; they depend on the solution of the
preferred-basis problem. As shown in Sec.\ \ref{sec:preferred_basis}, this
problem can be formulated as a highly nontrivial variation problem known as the
predictability sieve. Questions about the measurement of noncommuting
observables are also relevant only in the case of isolated subsystems, which
have not yet been included in the theory. This question will therefore be
revisited in Sec.\ \ref{sec:freedom_of_choice}, after the necessary concepts
have been developed.

\subsection{Historical remarks}

Most of the ideas described in Sec.\ \ref{sec:equivalence_classes} were
introduced by von Neumann \cite{vonNeumann1955secV4}. He discussed Eq.\
(\ref{eq:rho_hat}) extensively \cite{vonNeumann1955secV4}, but he always
expressed his reduction rule (\ref{eq:reduction_a}) in terms of Eq.\
(\ref{eq:rho_i_vN1}) rather than Eq.\ (\ref{eq:rho_i_vN2}). However, it seems
clear that von Neumann thought of himself as having formulated the latter rule
\footnote{On page 349 of Ref.\ \cite{vonNeumann1955}, following a discussion of
the ambiguity in the fine-grained reduction defined by Eqs.\
(\ref{eq:reduction_a}) and (\ref{eq:rho_i_vN1}) for degenerate observables, von
Neumann says that he ``shall succeed in V.4 in making some statements about the
result of a general measurement, on a thermodynamical basis''}. Perhaps he did
not write it down explicitly because he considered the concept of reduction to
become trivial in this case (cf.\ Jauch \cite{Jauch1964, Jauch1968}), due to the
identity $\eqcl{\rho} = \eqcl{\hat{\rho}}$.

Jauch also strongly emphasized the importance of equivalence classes for the
measurement problem \cite{Jauch1964, Jauch1968}. However, he used the L\"uders
reduction instead of the vN--DLP reduction, which corresponds to a different
kind of equivalence class \cite{Foreman2019b}. It is not clear whether this was
a deliberate choice, because Jauch did not mention von Neumann's prior work in
this area. This is somewhat surprising given that Jauch has acknowledged that
much of his research is an ``elaboration of [the] work of von Neumann'' \cite{[]
[{, p.\ 299.}] Jammer1974}.

From a modern perspective, the main element missing from Jauch and von Neumann's
work on equivalence classes is a connection with the concepts of decoherence
theory. Integrating these concepts into the theory leads to a recognition that
equivalence classes are contingent, dynamically evolving entities and that
integrated and isolated subsystems must be treated differently (see Sec.\
\ref{sec:partial_reduction}). Also, neither Jauch nor von Neumann used the
criterion of dynamical stability (cf.\ Sec.\ \ref{sec:dynamical_stability}) to
motivate their introduction of equivalence classes.

Daneri, Loinger, and Prosperi \cite{Daneri1962} used ergodic theory to derive
the reduced state (\ref{eq:rho_i_vN2}) from the time-averaged dynamics of a
microscopic system interacting with a macroscopic apparatus (see also Prosperi
\cite{Prosperi1971, Prosperi1974}). However, in the present theory, Eq.\
(\ref{eq:rho_i_vN2}) is not used to describe isolated subsystems (see Sec.\
\ref{sec:partial_reduction}).

\section{Emergence of statistical frequencies}

\label{sec:emergent_statistics}

In the previous section, it was noted that, for the GMH demon, the outcome of an
experiment is to be identified with the manifold $\mathcal{M}_i$ generated during
the reduction process, and the comparison between theory and experiment is
mediated entirely by the predicted set of probabilities $\{ w_i \}$.  However, 
up to this point, no connection between individual outcomes and probabilities 
has yet been established.  The transition between these two modes of description
involves another level of emergence.

To describe this type of emergence, one must introduce a qualitatively new
concept, that of the ensemble. This is most conveniently done by working with
histories (see Sec.\ \ref{sec:emergent_noninterference}) and defining an
ensemble of subhistories. Each subhistory $h$ is defined with reference to a
particular subsystem and covers a finite interval of time. The subhistories can
be arranged in parallel (different subsystems during the same time interval) or
in series (the same subsystem during different time intervals) or any
combination of the two.

This physical ensemble is not yet a statistical ensemble, because its
subhistories are not independent. Each subsystem in the parallel
ensemble, for example, belongs to the environment of all the others.
Independence of the ensemble elements can only be achieved if we declare the
interactions between them to be irrelevant; this is put into practice by, in part, 
coarse-graining over the environment of each subsystem. Ignoring such
differences between subhistories is the key to converting a physical ensemble
into a (finite) statistical ensemble.

The definition of environments is therefore essential to the construction of
ensembles and, consequently, to the gathering of information by ``information
gathering and utilizing systems.'' It is thus doubtful whether it is possible to
formulate a self-consistent theory of such systems in which the concept of
environment is dismissed on the grounds that it is ``artificial or poorly
defined'' \cite{GellMannHartle1994a}.

From this point on, the arguments leading to the emergence of statistical
frequencies are well known \cite{Everett1957, Finkelstein1963, Hartle1968,
Graham1973, FarhiGoldstoneGutmann1989, CavesSchack2005, Omnes1999a_probability}.
For example, one can define a frequency operator for the ensemble
\cite{Finkelstein1963, Hartle1968, Graham1973, FarhiGoldstoneGutmann1989,
CavesSchack2005}, the eigenvalues of which become ever more narrowly distributed
around the Born-rule subhistory probabilities $w_h$ as the number of ensemble
elements increases. In the limit of an infinite ensemble, we can say that $w_h$
emerges as a well-defined statistical frequency. The rules of classical
probability theory can then be used to compare these values with the
probabilities $w_i$ in Eq.\ (\ref{eq:weight}).

But in order for this to have any meaning in a real (finite) ensemble, we must
agree to ignore all ``maverick'' histories in which the observed frequencies
deviate substantially from the norm. Such a choice is fundamental to the concept
of probability, and it cannot be derived from anything else
\cite{CavesSchack2005, Geroch1984}. That is, it represents a true case of
emergence.

Note that the statistical frequencies of the ensemble can only be meaningfully
compared with the subhistory probabilities $\{ w_h \}$ if the projector sets
$\mathcal{P}_h$ are the same for each subhistory in the ensemble. The limitation
to a single set of projectors discussed in Sec.\
\ref{sec:incomplete_information} for the case of a closed system is therefore
also applicable to the case of an ensemble of open integrated subsystems.

\section{Partial reductions and isolated subsystems}

\label{sec:partial_reduction}

Let us now examine the consequences of relaxing the restriction to integrated
subsystems imposed at the end of Sec.\ \ref{sec:integrated_isolated}. The
purpose of that restriction was to circumvent difficulties arising from widely
divergent decoherence times $\tau_{\text{dec}} (k)$ for different subsystems
$k$. If these decoherence times differ greatly, it becomes necessary to consider
reductions involving only a few subsystems at a time, rather than the reduction
of all subsystems at once. Subsystems to be reduced are placed in a set $A$,
while all other subsystems belong to the complementary set $B$.

\subsection{Partial reductions}

A partial reduction of the subsystems in set $A$ can be generated by choosing
all projectors (\ref{eq:compound_projector}) in the set (\ref{eq:P_set}) such
that for all $k \in B$, the set $\{ \tilde{P}_{k i_k} \}$ contains only the
single element $1_k$. Such projectors then take the form
\begin{equation}
P_{i} = \prod_{k} P_{k i_k} = \prod_{k \in A} P_{k i_k} = P_{i}^A \otimes 1_{B}
, \label{eq:P_i_partial}
\end{equation}
in which $P_i^A$ acts only in $A$. The decoherence time
(\ref{eq:decoherence_max}) is likewise replaced with that of set $A$:
\begin{equation}
\tau_{\text{dec}} (A) = \max_{k \in A} \tau_{\text{dec}} (k) .
\end{equation}
Note, however, that simply using the projector (\ref{eq:P_i_partial}) in the
theory of Sec.\ \ref{sec:emergence_outcomes} does not yield the desired results,
because this would generate a complete reduction in which the subsystem states
$k \in B$ are all reduced to $1_k$. A separate theory of partial reductions is
therefore necessary.

The root of the problem can be seen by noting what would happen if we simply use
the set (\ref{eq:P_set}) in the equivalence class
(\ref{eq:equivalence_compound}). All members of a given equivalence class
$\eqcl{\rho}$ would then have identical probabilities (\ref{eq:weight}).
However, this criterion alone is insufficient to define the concept of a partial
reduction, because it does not impose any constraint on the subsystems in $B$. A
partial reduction that is applied only to the subsystems in $A$ should minimize
changes in the subsystems in $B$. This can be achieved by imposing a stronger
constraint in which membership in $\eqcl{\rho}$ is defined also by the values of
the operators $\tr_A (\rho P_{i})$. Here $\tr_A \sigma$ denotes a partial trace
of $\sigma$ over the subsystems in $A$, the result of which is an operator in
$B$.

\subsection{Equivalence class and reduction process}

\label{sec:partial_reduction_equivalence_class}

With this in mind, the equivalence class of $\rho$ for a partial reduction
generated by the set (\ref{eq:P_set}) can be defined as [cf.\ Eq.\
(\ref{eq:equivalence_compound})]
\begin{equation}
\eqcl{\rho} = \{ \sigma : \tr_A (\sigma P_{i}) = 
\tr_A (\rho P_{i}) \ \forall P_i \in \mathcal{P} \} . \label{eq:eqcl_partial}
\end{equation}
As before, the quantum state is identified with $\eqcl{\rho}$. The associated
reduced state (\ref{eq:rho_hat}) can then be correspondingly redefined as
\begin{subequations}
\label{eq:rho_hat_new}
\begin{gather}
\hat{\rho} = \sum_i w_i \rho_i , \qquad \rho_i = \rho_i^A \otimes \rho_i^B , \\
\rho_i^A = \frac{P_i^A}{d_i^A}  , \qquad
\rho_i^B =  \frac{\tr_A (P_i \rho)}{w_i} \quad (w_i \ne 0) ,
\end{gather}
\end{subequations}
in which $d_i^A = \tr P_i^A$. Here $\rho_i^A$ and $\rho_i^B$ are normalized
density operators belonging to $A$ and $B$, respectively. The former represents
a direct reduction of the $A$ component of $\rho$ [cf.\ Eq.\
(\ref{eq:rho_i_vN2})], whereas the latter represents an indirect reduction of
the $B$ component of $\rho$. This indirect reduction is a consequence of the
correlations between $A$ and $B$ generated by their interactions.

The concept of indirect reduction of a microscopic subsystem by a macroscopic
apparatus was elucidated by Landau and Lifshitz \cite{[] [{, \S 7.}]
LandauLifshitz1977}. Their reduction process is formulated in terms of the
relative-state expansion for pure states popularized by Everett
\cite{Everett1957, Everett1973}. The expression for $\hat{\rho}$ in Eq.\
(\ref{eq:rho_hat_new}) is similar but not identical to a relative-state
expansion. The density operator $\rho_{i}^{B}$ is called the \emph{conditional
state} of $B$ given the state $\rho_{i}^{A}$ of $A$ \cite{[] [{, p.\ 433.}]
SchumacherWestmoreland2010}. Such conditional states play an important role in
the definition of quantum discord \cite{Zurek2000, HendersonVedral2001,
OllivierZurek2001, ModiVedral2012}.

As shown in Appendixes \ref{app:eqcl_representation} and
\ref{app:partial_reduction}, the definition (\ref{eq:rho_hat_new}) of
$\hat{\rho}$ shares many of the properties of the reduced state
(\ref{eq:rho_hat}) described in Sec.\ \ref{sec:equivalence_classes}. The most
important of these are Eqs.\ (\ref{eq:rho_hat_eqcl}),
(\ref{eq:eqcl_representation}), and (\ref{eq:maximum_entropy}), which establish
that $\eqcl{\rho}$ can be represented mathematically by $\hat{\rho}$.

One may wonder whether the definition (\ref{eq:eqcl_partial}) contradicts the
justification given for the introduction of equivalence classes in Sec.\
\ref{sec:equivalence_classes}, because membership in $\eqcl{\rho}$ is no longer
defined exclusively by empirically meaningful probabilities \footnote{The
equivalence class (\ref{eq:eqcl_partial}) can be expressed in terms of
\emph{counterfactual} probabilities for reductions that are not actually
performed \cite{Foreman2019b}. However, such a description is not given here.}.
However, the theory of Sec.\ \ref{sec:emergence_outcomes} is not (and cannot be)
formulated entirely in terms of such probabilities, because the state $\rho$ is
required to follow unitary dynamics (\ref{eq:unitary}) during the intervals
between reductions. The additional constraints on equivalence-class membership
imposed in Eq.\ (\ref{eq:eqcl_partial}) merely require the $B$ component of
$\rho$ to adhere as closely as possible to this unitary dynamics---complete
adherence being impossible due to the effect of indirect reduction.

Given that $\eqcl{\rho}$ is no longer defined solely by probabilities, no simple
expression of the type shown in Eq.\ (\ref{eq:Delta_eqcl}) can be given for the
distance between equivalence classes.  However, this distance can still
be defined as the trace distance between the class representatives
[cf.\ Eq.\ (\ref{eq:distance_rho_hat})]:
\begin{equation}
\Delta (\eqcl{\rho}, \eqcl{\rho'}) = D (\hat{\rho}, \hat{\rho}') .
\end{equation}
The vanishing of the distance between $\eqcl{\rho}$ and $\eqcl{\hat{\rho}}$
shown in Eq.\ (\ref{eq:distance_zero}) remains valid, of course, because
$\eqcl{\hat{\rho}} = \eqcl{\rho}$. All of the qualitative conclusions in Sec.\
\ref{sec:emergence_outcomes} regarding the emergence of outcomes can therefore
be carried over to the case of partial reductions.

\subsection{Isolated subsystems}

\label{sec:isolated_subsystems}

An isolated subsystem (cf.\ Sec.\ \ref{sec:integrated_isolated}) may be
defined as one that effectively never decoheres:
\begin{equation}
\tau_{\text{dec}} (k) \to \infty .
\end{equation}
Such subsystems are permanently excluded from $A$. Their states are therefore
reduced only indirectly \cite{LandauLifshitz1977}, most commonly during
measurement situations in which the state of the subsystem becomes strongly
correlated with the state of a macroscopic measuring apparatus. The
recognition that an indirect reduction is as good as a direct reduction
\cite{EPR1935, LondonBauer1939t} played a significant role in the history of
quantum mechanics, leading Bohr to abandon the local ``disturbance'' concept
previously associated with measurement \cite{Bohr1935}. This was also the
motivation for the use of the conditional states $\rho_i^B$ in the definition of
quantum discord \cite{Zurek2000, HendersonVedral2001, OllivierZurek2001,
ModiVedral2012}.

The apparatus can perform a measurement only if it is an integrated subsystem
\cite{PeresZurek1982, Peres1984b, Peres1986, Peres1988b}, because this is what
allows it to undergo the decoherence necessary for the emergence of definite
outcomes \cite{Anderson1986}. This point was stressed by Heisenberg \cite{[] [{,
p.\ 57.}] Heisenberg1958a}: ``The measuring device deserves this name only if it
is in close contact with the rest of the world, if there is an interaction
between the device and the observer. \ldots If the measuring device would be
isolated from the rest of the world, it would be neither a measuring device nor
could it be described in the terms of classical physics at all.''

Although the direct reduction of isolated subsystems is not part of the basic
formalism of the present theory, such reduction processes can sometimes be
inserted ``by hand'' into a history without generating any interference
problems. However, these optional insertions can often be done in different ways
that are mutually inconsistent or complementary (i.e., the additional reductions
are consistent with noninterference individually but not jointly). This
possibility has led to much debate over the meaning of the histories formalism
\cite{dEspagnat1987b, dEspagnat1989b, dEspagnat1990, dEspagnat1995,
DowkerKent1996, BassiGhirardi1999, BassiGhirardi2000c, Omnes1992, Omnes1994,
Omnes1999a, Griffiths2002}. In response, Omn\`es \cite{Omnes1992, Omnes1994}
introduced a distinction between ``true'' events (which in the present context
are those arising from direct reduction of an integrated subsystem, including
the consequent indirect reduction of isolated subsystems) and ``reliable'' or
``trustworthy'' events (those arising from direct reduction of an isolated
subsystem). The latter have an inferior status because they are
subjective---they do not generate interference, but they are essentially
controlled by the whims of the theorist. This may be taken as justification for
the Copenhagen practice of ascribing meaning to the properties of isolated
subsystems only in the context of their interactions with a measuring apparatus.
In what follows, it is assumed that isolated subsystems are reduced only
indirectly.

Of course, even Landau and Lifshitz found it convenient to introduce a linear
operator in $B$ (generally \emph{not} a projection operator) to describe the
effect of indirect reduction on isolated subsystems \cite{LandauLifshitz1977}.
In the present case, such a description could be implemented using the quantum
operations formalism discussed in Sec.\ \ref{sec:non_projective_measurements}.
However, this is only an alternative mathematical way of expressing the same
physics of information acquisition. At the fundamental level, the reduction of
isolated subsystems remains an indirect process.

\subsection{Partially isolated subsystems}

\label{sec:partially_isolated_subsystems}

Thus far we have considered explicitly only integrated subsystems, which are
directly reduced at every reduction time $t_j$, and isolated subsystems, which
are never directly reduced. Of course, the theory in Sec.\
\ref{sec:partial_reduction_equivalence_class} can also be applied to partially
isolated subsystems, which are directly reduced occasionally but not always.
However, this case brings in no qualitatively new features, so it is not
discussed further here.

\subsection{Quantum-state tomography}

\label{sec:quantum_state_tomography}

In Sec.\ \ref{sec:incomplete_information}, it was noted that the usual textbook
description of quantum-state tomography \cite{dEspagnat1976, GottfriedYan2003},
in which the density operator $\rho$ is determined by ``measuring'' the value of
$\tr (P \rho)$ for a complete set of projectors $P$, is not applicable to the
case of a closed system, nor can it be used to find the density operator of an
open integrated subsystem. The reason for this is that the only projectors that
can be ``measured'' in such cases are those belonging to the orthogonal set that
defines the reduction that actually occurs at a given time. Such an orthogonal
set is never complete in the relevant sense, because it defines only an
equivalence class of density operators, not a unique density operator.

The definition of a unique density operator by means of quantum-state tomography
can, however, be achieved for an assembly of open \emph{isolated} subsystems
through the use of \emph{indirect} reductions. The state is first prepared by
performing an indirect reduction generated by some given measuring apparatus.
An assembly of identically prepared subsystems is then allowed to interact with
various other types of apparatus. If the indirect reductions generated by these
apparatuses can be formally described as equivalent to direct reductions that
form a complete set of operators, the density operator of the isolated subsystem
is thereby defined uniquely.

Note that the assembly considered here is not an ensemble in the sense defined
in Sec.\ \ref{sec:emergent_statistics}. An assembly of identically prepared
isolated subsystems interacting with \emph{identical} apparatuses forms an
ensemble, because its statistical frequencies can be meaningfully compared with
theoretical probabilities. However, no such direct comparison is possible for an
assembly of isolated subsystems interacting with different apparatuses. Such an
assembly is, rather, a collection of different (in general complementary)
statistical ensembles.

\subsection{Consistency of secondary measurements}

\label{sec:secondary_measurements}

d'Espagnat \cite{dEspagnat1976_ch15_16} has criticized Jauch's \cite{Jauch1964}
use of equivalence classes to describe an isolated subsystem interacting with a
measuring apparatus on the grounds that secondary measurements performed by
another apparatus for the purpose of distinguishing between the members of such
an equivalence class are difficult but not fundamentally impossible. According
to this argument, one cannot say that the description of experimental outcomes
as equivalence classes is objective even in a weak sense, because an
inconsistency in the theoretical description could be demonstrated if a
secondary measurement of this type is performed. Bub \cite{Bub1968a} has also
criticized the theory of Daneri, Loinger, and Prosperi \cite{Daneri1962} on
similar grounds.

A rejoinder to such anticipated criticism was given already by Jauch
\cite{Jauch1964} on the basis of Bohr's oft-repeated statement that the very
possibility of measurement implies a classical apparatus. Although there is a
kernel of truth in this statement, such a terse reply is less than fully
satisfying. Here this question is examined from the perspective of decoherence
theory, which provides a more complete answer.

The basic issue is that in order to demonstrate that the isolated subsystem and
primary apparatus are in an entangled state rather than a statistical mixture,
it is necessary to isolate this composite system so that it does not become
entangled with its environment. Assuming that this can be done, the composite
system does not decohere; it therefore plays the same role as the isolated
subsystem in Sec.\ \ref{sec:isolated_subsystems}. That is, it is reduced only
indirectly, as a consequence of its interaction with the secondary apparatus.
But the isolation of the primary apparatus from the outside world means that it
no longer functions as a measuring device \cite{PeresZurek1982, Peres1984b,
Peres1986, Peres1988b} (see the quotation from Heisenberg in Sec.\
\ref{sec:isolated_subsystems}). The secondary apparatus is, of course, presumed
to be integrated, so that it can decohere and perform its function as a
measuring device.

Therefore, the relevant point is not that measuring non-collective variables of
a measuring apparatus is difficult, but rather that doing so requires an
entirely different experimental arrangement. Even if future advances in
technology allow such demanding experiments to be performed, this would have no
bearing on the question of whether the original experiment was consistently
described. The decisive concept is once again that ``unperformed experiments
have no results'' \cite{Bohr1935, Peres1978}.

\subsection{Classical and quantum measurements}

\label{sec:classical_quantum_measurements}

The basic issue can be restated concisely using the concepts of classical and
quantum measurements. A classical measurement is what we infer from the
correlations between the states of two integrated subsystems. A quantum
measurement is what we infer from the correlations between the states of an
isolated subsystem and an integrated subsystem.

Any number of classical measurements can be concatenated without changing the
experimental arrangement. But a quantum measurement of a quantum measurement is
a contradiction in terms, because a measuring apparatus cannot be isolated by
definition. Attempting to isolate a quantum measuring apparatus so that it can
be subjected to a quantum measurement changes the entire experimental
arrangement. The modified experimental arrangement is therefore irrelevant to
the question of whether the original experiment was consistently described.

%Wigner's friend \cite{PeresZurek1982, FuchsMerminSchack2014, Brukner2017}

\section{The preferred-basis problem}

\label{sec:preferred_basis}

The preferred-basis problem was set aside at the end of Sec.\
\ref{sec:dynamical_stability} in order to focus on the problem of outcomes. Let
us now reexamine the former problem in the light of what has been learned since
then. The goal is to find a way of dealing with the preferred-basis problem that
is consistent with the concept of dynamical stability introduced in Sec.\
\ref{sec:dynamical_stability}.

\subsection{The predictability sieve}

\label{sec:predictability_sieve}

Zurek has developed a variational method for this purpose known as the
``predictability sieve'' \cite{Zurek1993a, ZurekHabibPaz1993, Zurek1998,
Zurek2003}, in which projectors at an initial time $t_a$ are chosen so as to
yield the most predictable results (in a sense to be described below) at some
later time $t_b = t_a + \Delta t$. This section describes a modified version of
the predictability sieve that is better suited for use with the reduction
process defined by Eqs.\ (\ref{eq:reduction}) and (\ref{eq:rho_hat_new}). Note
that the times $t_a$ and $t_b$ used here are different from those in Secs.\
\ref{sec:macroscopically_local} and \ref{sec:dynamical_stability} \footnote{The
time $t_a$ in Sec.\ \ref{sec:predictability_sieve} is equivalent to the time
$t_b$ in Secs.\ \ref{sec:macroscopically_local} and
\ref{sec:dynamical_stability}.}.

The basic idea of the modified predictability sieve is quite simple. Let $\rho
(t_a)$ be the density operator at time $t_a$. At this time, we can use some
given set of projectors (\ref{eq:P_set}) to define the reduced states $\rho_{i}
(t_a)$ in Eq.\ (\ref{eq:rho_hat_new}). Let these states evolve unitarily until
the final time $t_b$, yielding
\begin{equation}
\rho_{i} (t_b) = U (t_b, t_a) \rho_{i} (t_a) U (t_a, t_b) .
\label{eq:unitary_ab}
\end{equation}
At this time, the first stage (\ref{eq:reduction_a}) of a second
reduction is performed using the same projector set $\mathcal{P}$,
yielding the reduced states $\hat{\rho}_{i} (t_b)$.

The modified predictability sieve requires us to choose $\mathcal{P}$
so as to minimize the functional
\begin{subequations}
\label{eq:predictability_functional}
\begin{equation}
G = \sum_{i} w_{i} (t_a) \{ S [ \hat{\rho}_{i} (t_b) ] -
S [ \rho_{i} (t_a) ] \} , \label{eq:predictability_functional_a}
\end{equation}
the value of which is the mean entropy generated after the first reduction. This
is the same as the mean entropy generated by the second reduction:
\begin{equation}
G = \sum_{i} w_{i} (t_a) \{ S [ \hat{\rho}_{i} (t_b) ] -
S [ \rho_{i} (t_b) ] \} , \label{eq:predictability_functional_b}
\end{equation}
\end{subequations}
because the unitary evolution (\ref{eq:unitary_ab}) does not change the entropy.
This functional therefore satisfies $G \ge 0$, due to the inequality
(\ref{eq:increased_entropy}). 

Note that the entropy generated within each manifold $\mathcal{M}_i^A$ (defined
by the projector $P_i^A$) has already been maximized as a consequence of the
definition of the reduced state (\ref{eq:rho_hat_new}). Since a change of states
within $\mathcal{M}_i^A$ is regarded as an irrelevant change of internal
variables, the minimization of $G$ can be viewed as a minimization of the
remaining \emph{macroscopic} entropy generated by any changes in the
\emph{collective} variables.

Why is this called a predictability sieve? Minimization of the
entropy-generation functional (\ref{eq:predictability_functional}) means that
the nonvanishing conditional probabilities for the second reduction are
concentrated on the smallest possible number of manifolds $\mathcal{M}_j^A$; the
evolution of the states $\rho_i (t)$ after the first reduction is therefore as
close to deterministic as possible.

This has the side effect of maximizing correlations between subsystems, because
uncorrelated subsystems would have probabilities spread over more manifolds
$\mathcal{M}_j^A$. The predictability sieve therefore tends to generate a
\emph{redundancy of records} in different subsystems, such that knowledge of the
collective variables in one subsystem provides information about the collective
variables in other subsystems. The idea that a form of weak objectivity emerges
from the proliferation of such records is known as quantum Darwinism
\cite{OllivierPoulinZurek2004, OllivierPoulinZurek2005, BlumeKohoutZurek2006,
Zurek2007b, Zurek2009, RiedelZurekZwolak2012, Zurek2018a}.

The predictability sieve consequently maximizes predictability in two distinct
ways.  Not only does it select reduced states whose behavior is as close to
deterministic as possible; it also allows the properties of subsystems
to be inferred from information about other subsystems.

This version of the predictability sieve uses the same general
entropy-minimization concept proposed in Zurek's original work \cite{Zurek1993a,
ZurekHabibPaz1993, Zurek1998, Zurek2003}. It differs, however, in the details of
how the entropy is defined. The original sieve dealt with individual pure states
of a given subsystem and generated entropy by taking a partial trace over the
environment of that subsystem. In contrast, the functional
(\ref{eq:predictability_functional}) is formulated directly in terms of the
entropy generated by the reduction process defined by Eqs.\
(\ref{eq:reduction}) and (\ref{eq:rho_hat_new}). It therefore deals with all
subsystems together and with the entire set of projectors (\ref{eq:P_set}) as a
whole.

The main conceptual difference lies in the choice of what is defined to be
irrelevant in order to generate entropy. One irrelevant quantity here is the
distinction between different states in a manifold $\mathcal{M}_i^A$. As
described in Sec.\ \ref{sec:microstates_macrostates}, this concept of
irrelevance cannot always be handled by decomposing subsystems into a tensor
product of a collective subsystem and an internal environment. The recipe of
tracing over an internal environment is therefore generally incapable of
handling the relevance concept associated with collective variables.

Tracing over an environment also does not mean that the environment is treated
as irrelevant, at least in the context of Zurek's formulation of the
predictability sieve. For example, if the total system is in a pure state, the
entropy of any subsystem is the same as the entropy of its environment
\cite{vonNeumann1955, NielsenChuang2000}, so minimizing the former is the same
as minimizing the latter. In any event, treating an external environment as
truly irrelevant would be inappropriate even in cosmological contexts, because
the state of an entirely unknown environment should for consistency be described
as a state of maximum entropy. But this corresponds to an infinite-temperature
environment, which conflicts with the experimental evidence showing that our
cosmological environment is better described as a heat bath at a temperature of
about 2.7~K.

Despite the difference in detail between the definitions of the two sieves, the
main qualitative conclusions derived by applying Zurek's original sieve to
simple model systems are likely to hold for the modified sieve as well. In
particular, note that neither version of the predictability sieve makes any
reference to a criterion of macroscopic locality in coordinate space (cf.\ Sec.\
\ref{sec:macroscopically_local}). The decoherence of macroscopically nonlocal
superpositions is, rather, presumably a contingent dynamical \emph{consequence}
of the predictability sieve for many typical initial states $\rho (t_a)$ and
Hamiltonians $H$ and does not need to be imposed as an axiom. No explicit tests
of this statement are performed here, however.

\subsection{Constrained variation and dynamical stability}

In the above definition of the predictability sieve, the projector set
$\mathcal{P}$ was assumed not to change during the time interval $\Delta t$.
This does not mean that $\mathcal{P}$ is independent of time, because the
application of the sieve at a succession of different times $t_a$ will give a
succession of different values of $\mathcal{P}$. What it does mean is that this
predictability sieve is applicable only to projector sets that are slowly
varying functions of time. That is, it must be possible to choose the interval
$\Delta t$ between reductions to satisfy Eq.\ (\ref{eq:slow_time}), in which
$\tau_{\mathcal{P}}$ is the timescale for the dynamics of $\mathcal{P}$. But
this is just the condition for dynamical stability introduced in Sec.\
\ref{sec:dynamical_stability}.

The constraint $\mathcal{P} (t_b) = \mathcal{P} (t_a)$ could presumably be
relaxed by allowing $\mathcal{P} (t_b)$ to vary independently of $\mathcal{P}
(t_a)$ and searching for minima of $G$ in the neighborhood of those derived from
this constraint. However, the assumption that $\mathcal{P} (t_b) \approx
\mathcal{P} (t_a)$ could not be eliminated without violating the criterion of
dynamical stability.

\subsection{Variation of subsystem decomposition}

Note that the projector set $\mathcal{P}$ depends on the set of variables
$\mathcal{S}$ used to define the subsystem decomposition: $\mathcal{P} =
\mathcal{P} (\mathcal{S})$.  The variation of $\mathcal{P}$ in the search
for minima of $G$ therefore generally involves variation of $\mathcal{S}$ as
well.  In this way, the predictability sieve provides a criterion for defining 
the time dependence of subsystem decompositions, at least in regard to 
macroscopic subsystems.

The variations of $\mathcal{S}$ and $\mathcal{P}$ may in general have to be
performed separately. For example, in typical quantum measurement situations, an
isolated subsystem interacts with a macroscopic measuring apparatus that is
initially in a metastable state. During this interaction there is a significant
probability of triggering a phase transition, such as the condensation of a
liquid droplet in a cloud chamber. For most practical experiments, the
description of such amplification processes involving metastable states would
require the use of thermodynamic variables such as temperature. This introduces
an additional layer of complication into the theory, because the subsystems
would then have to be defined at the thermodynamic level, as discussed in Sec.\
\ref{sec:emergence_meaning}. That is, thermodynamic concepts would emerge as
usual from a conceptual statistical ensemble of quantum states in, say, a
quasi-canonical distribution. For the individual members of the ensemble, the
subsystem decompositions $\mathcal{S}$ would be treated as \emph{fixed}, and the
predictability sieve would be applied only to the projector sets $\mathcal{P}$.
Variation of $\mathcal{S}$ would be performed in a separate predictability sieve
applied to the ensemble as a whole. However, to avoid inessential complications,
this thermodynamic type of description is not considered further here.

\subsection{Restriction of freedom to choose subsystems}

The predictability sieve imposes a very powerful restriction on the freedom to
choose the subsystems $\mathcal{S}$ and projector sets $\mathcal{P}$. Choices
that violate these restrictions are not logically forbidden, but they are
basically useless for any type of practical calculations or meaningful
comparison with experiment. As noted by Bohr \cite{Bohr1935}, a limited degree
of freedom remains, due in this case to the presumed existence of multiple
(possibly infinitely many) viable solutions of the variation problem. However,
this restricted set of choices is, by any reasonable criterion, of measure zero
in comparison to the completely arbitrary set that describes the possibilities
prior to the solution of the preferred-basis problem. The remaining freedom of
choice has no observable consequences as long as the projectors $P_i$ and the
reduction intervals $\Delta t$ are fine-grained enough to be well below the
resolution limit of any protoscientists using the theory. In this way, the
seemingly arbitrary choice of $\mathcal{S}$ and $\mathcal{P}$ becomes weakly
objective rather than subjective.

\subsection{Choice of microscopic observables}

\label{sec:freedom_of_choice}

At this point we are in a position to address the question raised at the end of
Sec.\ \ref{sec:meaning_of_results}. Namely, is it not a trivialization of the
measurement problem if we only deal with orthogonal sets of projectors at any
given instant of time? The questions that troubled the founders were related,
rather, to the freedom of the experimenter to choose any observable from a
noncommuting set. Indeed, the presumption that the experimenter has the freedom
to control the conditions of an experiment is a necessary prerequisite for any
pragmatic account of experimental physics \cite{Bohr1935, PeresZurek1982,
Peres1986c, Peres1995, Primas1990a0}.

This freedom has already implicitly been incorporated into the definition of the
assembly of isolated subsystems used to perform quantum-state tomography in
Sec.\ \ref{sec:quantum_state_tomography}. The basic issue is that the choice of
observable is determined by the choice of measuring apparatus used to achieve
the indirect reduction of the state of the isolated subsystem. However, as shown
above, the emergence of outcomes occurs on the timescale $\Delta t \ll
\tau_{\mathcal{P}}$ defined in Eq.\ (\ref{eq:slow_time}). But questions about
the free will of the experimenter become meaningful only in the opposite limit
of time intervals comparable to or greater than $\tau_{\mathcal{P}}$, where
other levels of emergence come into play.

The future dynamics of planets in the solar system can be predicted with near
certainty even for time intervals much greater than $\tau_{\mathcal{P}}$.
However, the future of individual human beings (and their proxies, such as the
pseudo-random switches in the Aspect experiment \cite{Aspect1982b}) is
predictable only for time intervals close to $\tau_{\mathcal{P}}$, beyond which
the dynamics quickly becomes almost entirely unpredictable. On these longer
timescales, one can therefore invoke the free will of the experimenter without
fear of contradiction, even though this concept is nowhere to be found among the
basic principles of quantum mechanics. It should be stressed that this level of
emergence does not require the formulation of any model for the consciousness of
the experimenter; it depends only on the assumption that the choices of the
experimenter are wholly unpredictable and can thus be treated as ``free''
variables within the theory \cite{Peres1986c, Peres1995}.

To put this in concrete terms, if an experimenter has set up a given experiment,
has placed her finger on the ``start'' button, and the nerve impulses to press
the button are already racing down her arm, she certainly does not have the
freedom to swap this experiment for another. That freedom comes into being only
on longer timescales $\tau_{\mathrm{fw}} \gg \tau_{\mathcal{P}}$, when her
actions are no longer predictable.

\subsection{Redundancy of records as a primary concept}

Riedel has recently suggested that the redundancy of records described in Sec.\
\ref{sec:predictability_sieve} can be used as a criterion for the definition of
branches of the quantum state vector \cite{Riedel2017}. This definition also
makes use of the condition of locality in coordinate space discussed in Sec.\
\ref{sec:macroscopically_local}, together with an explicit length scale defining
the spatial extent of nonlocality. It is not yet clear whether this length scale
can be eliminated from the definition \cite{Riedel2017}, but the concept of
locality in coordinate space appears to be fundamental to Riedel's method.

It is also unclear whether the redundancy of records can be developed into a
practical tool for the construction of branches that has all of the advantages
of the predictability sieve. Here the latter has been chosen because it is more
fully developed and the redundancy of records can be derived from it without the
need for any \emph{a priori} reference to locality in coordinate space.

One intriguing outcome of Riedel's work is a definition of compatibility of
recorded observables that does not require the corresponding operators to
commute \cite{Riedel2017}. It would be interesting to see whether this concept
can be developed further in a context free of locality constraints.

\section{Emergence of weak objectivity}

\label{sec:weak_objectivity}

\subsection{Weak objectivity in quantum mechanics}

As noted already in Sec.\ \ref{sec:ideal_observer}, the GMH demon is just a
useful idealization. All information gathered from experiments is gathered by
individual protoscientists; it is therefore unavoidably subjective. The
subjective origin of all quantum-mechanical information was acknowledged early
on by von Neumann \cite{vonNeumann1955}, London and Bauer
\cite{LondonBauer1939t}, and Wigner \cite{Wigner1967}, and this aspect of the
problem is embraced unflinchingly in the modern theory of quantum Bayesianism
(or QBism) \cite{CavesFuchsSchack2002a, Fuchs2010, Mermin2012a, FuchsSchack2013,
FuchsMerminSchack2014, Mermin2019}.

However, to avoid accusations of blatant subjectivity \cite{Daneri1962} or
solipsism \cite{Norsen2016}, it is helpful to formulate the theory in a way that
does not refer \emph{directly}, at least in the first instance, to the
limitations of humans or other individual observers. This is the role of the GMH
demon, for whom the relevant histories belong to a set roughly comparable to
what Gell-Mann and Hartle have called a ``maximal'' \cite{GellMannHartle1990,
Hartle1991a} or ``full'' \cite{GellMannHartle1991, GellMannHartle1993} set,
thereby defining a so-called quasiclassical realm \cite{GellMannHartle1994b,
GellMannHartle2007}. The word ``maximal'' suggests that this set should be as
fine-grained as possible, consistent with the limitations of decoherence and
predictability in Eq.\ (\ref{eq:slow_time}). It is unlikely that exact solutions
can be found for this variation problem, but for practical purposes all that is
needed is a set of histories that is much more fine-grained than the resolution
limits of any protoscientist using the theory. The histories relevant to these
protoscientists can then be taken to be coarse-grainings of the quasi-maximal
set defined for the GMH demon. Weak objectivity emerges in the limit in which
the fine-grained features of the demon's quasiclassical realm become
indiscernible to all protoscientists using the theory.

Of course, from this point of view, there is no guarantee that the
quasiclassical realm corresponding to our everyday experiences is unique
\cite{GellMannHartle1994b, Hartle2011}. However, since the existence of multiple
inequivalent quasiclassical realms would not affect the conclusion of weak
objectivity, this topic is not pursued further here.

\subsection{Strong objectivity?}

Many authors have tried to reformulate quantum mechanics as a strongly objective
theory. The temptation to reach for this prize is obvious, but it comes at a
price: one must forfeit any possible claims of weak objectivity, as these are
derived from emergence. If the attempt at strong objectivity fails, the theory
that remains is merely subjective.

In this genre, the class of theories known as dynamical-reduction models
\cite{BassiGhirardi2003, BassiLochan2013} is particularly well developed. In the
version known as continuous spontaneous localization, reduction is generated by
coarse-grained number operators rather than projection operators. The most
important difference from the present theory, however, is that the reduction
process is tethered to a fixed choice of preferred basis---namely, a
coarse-grained coordinate basis. The parameters of the model seem to be tailored
to mimic the emergence, in many common cases of decoherence, of such a basis as
the preferred basis \cite{JoosZeh2003}. But in decoherence theory, the preferred
basis is always contingent upon the form of the Hamiltonian and the initial
state $\rho$. As noted by Gell-Mann and Hartle \cite{GellMannHartle1990},
``there are no variables that are expected to decohere universally.''

This leads to a well-known difficulty, in that decoherence can occur by
mechanisms that do not generate a macroscopic displacement of particles in
coordinate space and thus do not immediately trigger a reduction in the
dynamical-reduction model \cite{AlbertVaidman1989}. In responding to this
criticism, Ghirardi \emph{et al.}\ have argued that ultimately, the detection of
any measurement outcome in the eyes and nervous system of a human observer would
definitely trigger such a reduction \cite{Aicardi1991,
GhirardiGrassiBenatti1995, BassiGhirardi2003}. They have also pointed out that
such a process does not imply any role for the consciousness of a sentient
observer.

Nevertheless, the mere presence of such discrepancies shows that a
coarse-grained coordinate basis cannot serve as the foundation for a weakly
objective theory. They also raise the suspicion that such a basis would be an
inadequate foundation for a strongly objective theory too. Ultimately the latter
question can only be decided by experiment. Dynamical-reduction models have
recently been subjected to increasingly stringent empirical tests, which so far
have not yielded any qualitative conclusions \cite{AdlerBassi2009,
BassiLochan2013, ArndtHornberger2014, Kovachy2015, VinanteBassi2016,
CarlessoBassi2016, Helou2017, VinanteBassi2017}.

The fundamental issue of concern here is closely related to the questions about
the concept of macroscopically local observables discussed in Secs.\
\ref{sec:macroscopically_local} and \ref{sec:local_observables_unobservable}.
However, there the criterion of macroscopic locality was used only in the
formulation of the pseudometric (\ref{eq:variation_distance}), rather than being
inserted directly into the definition of a preferred basis.

Similar comments can be made about pilot-wave theories \cite{BohmHiley1993,
Holland1993, DurrGoldsteinZanghi2013}. Here the ontology is based on particle
trajectories, even though the concept of a particle is known to be emergent
\cite{Davies1984, Wallace2008} and individual trajectories cannot be measured
\cite{Wiseman2007, Kocsis2011}, at least in the standard version of the theory
where particle probabilities are assumed to follow the Born rule exactly
\cite{Valentini2009}. Within this version of the theory, there are infinitely
many inequivalent ways of defining the trajectories, all of which yield the same
experimental predictions \cite{DeottoGhirardi1998}. All empirical predictions
are derived from the wave function itself and its decomposition into subsystems
\cite{[] [{; see Eqs.\ (5.17) and (6.1).}] DurrGoldsteinZanghi1992a}; the role
of the particle trajectories is simply to delineate a sequence of effective
reductions in the coordinate representation. At this level, the particle
trajectories are a purely subjective addition to the formalism of orthodox
quantum mechanics.

The only possible foundation for a claim of strong objectivity therefore lies in
the ``quantum non-equilibrium'' version of the theory in which the particle
trajectories do not satisfy the Born rule exactly \cite{Valentini2009}. Again,
claims of this sort can only be adjudicated experimentally. However, given that
both the concept of a particle and the special status of the coordinate
representation emerge from standard quantum mechanics, it is difficult to accept
such a proposal of strong objectivity in the absence of compelling evidence.

\section{Emergent reality}

\label{sec:emergent_reality}

The theory developed here shows that the reduction process can be regarded as
emerging from unitary dynamics under certain well-defined conditions. However,
this situation is unusual in science because the base-level theory (i.e., the
Schr\"odinger equation for a closed system) is not experimentally accessible
\cite{[] [{, p.\ 58.}] Heisenberg1930, [] [{; especially p.\ 27.}]
Heisenberg1955, Everett1973p98}. Only the overall theory comprising both
reduction and unitary dynamics can be tested in the laboratory. What does it
even mean to talk of ``emergence'' in this case?

It is also common to think of the base level as being more ``real'' than the
emergent level. The latter is sometimes called an ``illusion.'' However, the
word ``real'' is also usually restricted to what is empirically testable.  How
are we to deal with this situation?

The word ``reality'' has many different meanings. In one sense it refers to
something assumed to exist ``out there,'' independent of any observer. But since
we have no access to any such ``thing in itself,'' this concept is not very
useful on its own. It is more useful to talk about reality as an \emph{idea}
that we use to order and make sense of our experience \cite{[] [{; reprinted in }]
Einstein1936, *Einstein1950, *Einstein1954}. That is, we apply the
concept of reality to \emph{theories} rather than things.

One such idea is that of an absolute or totalitarian reality $R$, which is a
Boolean variable. A value $R \in \{ 0, 1 \}$ can always be assigned to any given
theory. However, this assignment must then be tested by comparing it with the
empirical or pragmatic reality $r$, which is a real variable in the range $0 < r
< 1$ (not $0 \le r \le 1$). It may be defined as
\begin{equation}
r = \exp (- \eta), \quad \eta = \max_{x \in \mathcal{D}} 
\frac{\abs{\Delta x}}{\delta x} ,
\end{equation}
in which $x$ is some set of variables, $\Delta x$ is the discrepancy between
theory and experiment, and $\delta x$ is a typical range of $x$ in some domain
of experience $\mathcal{D}$. If $\eta \ll 1$, we can say that the theory is
real in the pragmatic sense within this given domain. But such a result can
never strictly justify an assignment of $R = 1$ in the absolute sense.

A major difficulty with the concept of absolute reality is that any assignment
of $R = 1$ immediately becomes invalid whenever the theory is found to emerge
from another level. But this means that $R$ is a chimera, because (as noted in
the Introduction) there are no empirically validated theories (i.e., theories
with $\eta \ll 1$ in some domain) that are not already known to be emergent.
The only ``laws'' of physics thus far discovered are approximate ways of
describing some part of our experience, not clockwork mechanisms that control
the universe.

Hence, all we have left is the domain-dependent concept of pragmatic reality.
This has the advantage that the concept of reality now becomes meaningful in
many different domains; we no longer have to assume that everything we see is an
illusion. But we also no longer have the right to assign an absolute reality to
anything.

In this sense it is not meaningful to talk about the reality of the unitarily
evolving quantum state, prior to any concept of reduction. Pragmatic reality
deals only with the experimentally accessible combination of unitary evolution
and reduction. But the Schr\"odinger equation still plays an indispensable role
in this combination. As Feynman has observed \cite{[] [{, p.\ 2--9.}]
FeynmanLeightonSands1965v3}, ``it is not true that we can pursue science
completely by using only those concepts which are directly subject to
experiment.'' It is hard to imagine how a useful theory of open-system dynamics
could be formulated without making use at some point of assumptions about the
environment of the open system, developed in the context of a larger system
evolving according to the Schr\"odinger equation. And it is well known that
reducing the state prematurely will give results in conflict with experiment
\cite{Furry1936a}.

In recognition of this role, the unreduced quantum state is sometimes said to
represent a kind of ``veiled reality'' \cite{dEspagnat1995}. Weinberg expressed
this idea clearly \cite{[] [{, p.\ 79.}] Weinberg1992} when he said that ``wave
functions are real for the same reason that quarks and symmetries are---because
it is useful to include them in our theories.'' 

The quark is an apt analogy because the theory of quantum chromodynamics itself
predicts that free quarks can never be observed experimentally. Yet the concept
of a quark is now accepted as a part of reality because it is so useful
as an organizing principle for experiments in high-energy physics---in
particular, those performed in the domain of asymptotic freedom.

Likewise, Everett himself showed no compunction in acknowledging that
\cite{Everett1973p98} ``it is impossible for any observer to discover the total
state function of any physical system.'' Yet the total state of a closed system,
independent of any reduction, is useful enough to be recognized as a part of
reality, particularly in light of quantum-state tomography experiments used to
measure (in the statistical sense) the states of isolated subsystems---a type of
asymptotic freedom, as it were. Of course, emergent subsystems and the reduction
process are parts of this reality, too.

In a similar manner, it is meaningful to talk about the emergence of reduction
from unitary evolution simply because this concept is useful. Formulating the
measurement problem in terms of emergence is beneficial because it places both
the objective and subjective aspects of quantum mechanics front and center and
forces us to reconcile them. If nothing else, this approach may help to
short-circuit the heated debates over philosophical issues that tend to be
generated when objectivity is viewed as a binary concept, with nothing between
the two poles of strong objectivity and pure subjectivity.

\section{Conclusions}

\label{sec:conclusions}

Many authors have argued that the key to the measurement problem lies in the
macroscopic nature of the measuring apparatus and the corresponding inability
of an observer to distinguish among the many microstates consistent with a given
macrostate of the apparatus \cite{Daneri1962, Daneri1966, Weidlich1967,
VanKampen1988, VanKampen1991}. However, it has long been known that this
argument provides at best some partial support for a statistical interpretation
of the state vector; it does not justify the claim that either the state vector
or the density operator is consistent with the appearance of definite
experimental outcomes \cite{Wigner1963, JauchWignerYanase1967, Fine1970,
Bell1990, BassiGhirardi2000}.

Here I have shown that the problem of outcomes can be resolved, at least in the
sense of emergence, if the assertion of an observer's inability to distinguish
among these microstates is followed through and applied consistently in all
parts of the theory. Consistency demands that the quantum state be defined not
as a vector, ray, or density operator in Hilbert space but as an equivalence
class of density operators \cite{vonNeumann1955secV4, Jauch1964, Jauch1968}. For
integrated subsystems, such an equivalence class can be represented
mathematically by a projection operator \cite{vonNeumann1955secV4}, but if
isolated subsystems are involved, the class representative must also include the
associated conditional states. This definition must be applied every time the
quantum state is reduced, on a timescale slow in comparison to the decoherence
time but fast in comparison to the relevant quasiclassical dynamics. The
ubiquitous L\"uders reduction formula \cite{Lueders1950} retains too much
information. Trying to justify the appearance of distinct outcomes using this
formula is like trying to justify the second law of thermodynamics without
repeatedly discarding the irrelevant information generated during the mixing
process; it simply cannot be done.

The use of equivalence classes in the quantum theory of measurement has been
advocated by such luminaries as von Neumann \cite{vonNeumann1955secV4} and Jauch
\cite{Jauch1964, Jauch1968}. To the extent that their proposals are remembered
today, they seem to be regarded mainly as failed rivals to the throne currently
occupied by decoherence theory. However, as Zurek has pointed out
\cite{Zurek2018a}, a likely reason for the slow progress in almost a century of
work on the measurement problem is that its solution requires many ideas, not
just one. Here the definition of quantum states as equivalence classes is indeed
only one piece of the puzzle. Its utility is maximized by working in harmony
with the other concepts of decoherence theory, rather than as a competitor.

One of the most important of these concepts is that of dynamical stability. This
can be used to define a metric in which the distance between states before and
after reduction becomes arbitrarily small in the limit of slowly varying
projector sets, without referring to any criterion of locality in coordinate
space. This provides additional support for the introduction of equivalence
classes in which these states are defined to be identical.

The concept of dynamical stability also appears in the form of the
predictability sieve used to solve the preferred-basis problem. Here the
predictability sieve has been adapted to ensure its consistency with the
modified reduction process, but Zurek's basic principle of entropy minimization
\cite{Zurek1993a} remains the same. On the other hand, entropy is also
maximized, following Jaynes \cite{Jaynes1957a, *Jaynes1957b}, with respect to
the states within a given manifold defined by the collective variables. The
basic idea is to minimize the entropy of whatever we can, in principle, obtain
information about (i.e., the collective variables of integrated subsystems) and
maximize the entropy of everything else. This entropy maximization is at first
glance just a natural outcome of the definition of equivalence-class
representatives. However, at a deeper level it is seen to be a fundamental
postulate of dynamics, on par with the Schr\"odinger equation itself.

Another essential concept is that isolated subsystems are reduced only
indirectly, via their interaction with the directly reduced states of integrated
subsystems. The basis for this approach in standard quantum mechanics is
described clearly by Landau and Lifshitz \cite{LandauLifshitz1977}. In the
context of the present theory of equivalence classes, this indirect reduction of
isolated subsystems leads to the conditional states now widely used in the
definition of quantum discord \cite{Zurek2000, OllivierZurek2001,
HendersonVedral2001, ModiVedral2012}. The distinct roles played by isolated and
integrated subsystems in this theory are reminiscent of Bohr's insistence on the
``necessity of discriminating in each experimental arrangement between those
parts of the physical system considered which are to be treated as measuring
instruments and those which constitute the objects under investigation''
\cite{Bohr1935}. Bohr's argument, like the one given here, concerns the way in
which information is extracted from these subsystems.

In the resulting theory, the pristine beauty of the Schr\"odinger equation is
``continually abrogated by dice-miracles'' \cite{Schrodinger1953}. Nevertheless,
the existence of an emergent limit ensures that the abrogation is smooth enough
to be unnoticeable on some measures, thereby allowing a peaceful coexistence
between the objective features described by unitary dynamics and the subjective
features described by the reduction process. Because the simultaneous presence
of these objective and subjective features is mandated not just by quantum
mechanics but by the scientific method itself, it is unlikely that we can ever
``unscramble Bohr's omelette'' \cite{Jaynes1990} in the sense of isolating the
objective features in an empirically validated theory. Hence, we are stuck with
a \emph{weakly} objective theory that is good only for all practical purposes
\cite{Bell1990}. But that is good enough.

\begin{acknowledgments}

I am grateful to H. Dieter Zeh for critical comments on an earlier manuscript,
and to Enoch Yan Lok Tsui for discussions of the redundancy of records as a
criterion for the branching of quantum states.

\end{acknowledgments}

\appendix

\section{Fidelity and trace distance for L\"uders reduction of pure states}

\label{app:reduction_pure}

The trace distance (\ref{eq:trace_distance}) is bounded by the inequalities
\cite{FuchsGraaf1999}
\begin{equation}
1 - \sqrt{F (\rho, \sigma)} \le D (\rho, \sigma) \le \sqrt{1 - F (\rho, \sigma)} ,
\label{eq:trace_distance_bounds}
\end{equation}
in which the fidelity is defined as
\cite{NielsenChuang2000, Bengtsson2017}
\begin{equation}
F (\rho, \sigma) \equiv \Bigl( \tr \sqrt{\rho^{1/2} \sigma \rho^{1/2}} \Bigr)^2 
= F (\sigma, \rho) .
\label{eq:fidelity}
\end{equation}
If $\rho$ is a pure state, this takes the simple form \cite{NielsenChuang2000}
\begin{equation}
F (\rho, \sigma) = \tr \rho \sigma ,
\end{equation}
and the bounds (\ref{eq:trace_distance_bounds}) can 
be tightened to \cite{NielsenChuang2000}
\begin{equation}
1 - F (\rho, \sigma) \le D (\rho, \sigma) \le \sqrt{1 - F (\rho, \sigma)} .
\label{eq:trace_distance_bounds_tight}
\end{equation}
This is obtained by substituting $P = \rho$ into the inequality $\tr [P (\rho -
\sigma)] \le D (\rho, \sigma)$ [cf.\ Eqs.\ (\ref{eq:variation_distance}) and
(\ref{eq:trace_distance_inequality})].

The example discussed in Sec.\ \ref{sec:emergent_determinism} is the case of a
L\"uders reduction for a pure state $\rho$. The relevant fidelities are then
\begin{subequations}
\label{eq:D_simple}
\begin{align}
F (\rho, \hat{\rho}) & = \sum_i w_i^2 , \\
F (\rho, \rho_i) & = w_i ,
\end{align}
\end{subequations}
in which all quantities are evaluated at some given reduction time $t = t_j$.
Clearly, neither of the corresponding trace distances $D$ is small for the
general case in which more than one value of $w_i$ is significant.

If, however, the outcome $i = 1$ is dominant, it is convenient to write $w_1 = 1
- \epsilon$, where $\epsilon \ll 1$. For arbitrary $\epsilon$ we have $\sum_{i
\ne 1} w_i = \epsilon$, $0 \le w_{i \ne 1} \le \epsilon$, and $0 \le \sum_{i \ne
1} w_i^2 \le \epsilon^2$, hence
\begin{equation}
F (\rho, \hat{\rho}) = 1 - 2 \epsilon + \alpha \epsilon^2 \qquad 
(1 \le \alpha \le 2) .
\end{equation}
The relevant bounds (\ref{eq:trace_distance_bounds_tight}) on the trace distance
are
\begin{subequations}
\begin{align}
2 \epsilon - 2 \epsilon^2 & \le D(\rho, \hat{\rho}) \le \sqrt{2 \epsilon - \epsilon^2} , \\
\epsilon & \le D(\rho, \rho_1) \le \sqrt{\epsilon} , \\
1 - \epsilon & \le D(\rho, \rho_{i \ne 1}) \le 1 .
\end{align}
\end{subequations}
Thus, in the limit $\epsilon \to 0$, we have $D(\rho, \hat{\rho}) \to 0$,
$D(\rho, \rho_1) \to 0$, and $D(\rho, \rho_{i \ne 1}) \to 1$.
As noted in Sec.\ \ref{sec:emergent_determinism}, this leads
to the emergence of quasiclassical determinism.

\section{Isomorphism between equivalence classes and reduced states}

\label{app:eqcl_representation}

Here Eq.\ (\ref{eq:eqcl_representation}) is proved for the general partial
reduction discussed in Sec.\ \ref{sec:partial_reduction_equivalence_class}. The
changes involved in the special case of a total reduction (discussed in Sec.\
\ref{sec:equivalence_classes}) are indicated in the comments below Eq.\
(\ref{eq:sigma_hat}).

To prove Eq.\ (\ref{eq:eqcl_representation}), note that $\eqcl{\rho} =
\eqcl{\sigma} \Leftrightarrow \eqcl{\hat{\rho}} = \eqcl{\hat{\sigma}}$ is a
direct consequence of Eq.\ (\ref{eq:rho_hat_eqcl}), whereas $\hat{\rho} =
\hat{\sigma} \Rightarrow \eqcl{\hat{\rho}} = \eqcl{\hat{\sigma}}$ follows
immediately from the definition of an equivalence class. The only significant
part of Eq.\ (\ref{eq:eqcl_representation}) is thus the implication
$\eqcl{\hat{\rho}} = \eqcl{\hat{\sigma}} \Rightarrow \hat{\rho} = \hat{\sigma}$.
Here $\eqcl{\hat{\rho}} = \eqcl{\hat{\sigma}}$ means that
\begin{equation}
\tr_A [ (\hat{\rho} - \hat{\sigma}) P_{i}] = 0 \ \forall P_i \in \mathcal{P} ,
\label{eq:same_eqcl_partial}
\end{equation}
in which $\hat{\rho}$ and $\hat{\sigma}$ can be written as [cf.\ Eq.\
(\ref{eq:rho_hat_new})]
\begin{equation}
\hat{\rho} = \sum_{j} P_{j}^A \otimes R_{j}^{B} , \quad
\hat{\sigma} = \sum_{j} P_{j}^A \otimes S_{j}^{B} ,
\label{eq:sigma_hat}
\end{equation}
for some given sets of operators $\{ R_{j}^{B} \}$ and $\{ S_{j}^{B} \}$. (For a
total reduction, the set $A$ includes all subsystems, so the tensor product with
respect to $R_{j}^{B}$ reduces to multiplication by a number $R_j$.) The
quantity of interest is then
\begin{subequations}
\begin{align}
\tr_A [ (\hat{\rho} - \hat{\sigma}) P_i] & = \sum_{j} (R_{j}^{B}
- S_{j}^{B}) \tr (P_j^A P_i^A) \\
& = \sum_{j} (R_{j}^{B} - S_{j}^{B}) 
\delta_{ij} \tr P_j^A \\
& = (R_{i}^{B} - S_{i}^{B}) d_i^A .
\end{align}
\end{subequations}
Hence, from Eq.\ (\ref{eq:same_eqcl_partial}) we have
\begin{subequations}
\begin{align}
\eqcl{\hat{\rho}} = \eqcl{\hat{\sigma}} & \Rightarrow 
R_i^B = S_i^B \ \forall i \\ &
\Rightarrow \hat{\rho} = \hat{\sigma} ,
\end{align}
\end{subequations}
which was to be proved.

\section{Increase of entropy upon partial reductions}

\label{app:partial_reduction}

The equivalence class (\ref{eq:eqcl_partial}) and reduced state
(\ref{eq:rho_hat_new}) for the partial reduction discussed in Sec.\
\ref{sec:partial_reduction_equivalence_class} share many of the properties of
the equivalence class (\ref{eq:equivalence_compound}) and reduced state
(\ref{eq:rho_hat}) for the total reduction discussed in Sec.\
\ref{sec:equivalence_classes}. The most important of these properties are the
identities shown in Eqs.\ (\ref{eq:rho_hat_eqcl}),
(\ref{eq:eqcl_representation}), and (\ref{eq:maximum_entropy}).

The derivation of Eq.\ (\ref{eq:rho_hat_eqcl}) is a simple exercise that follows
immediately from the orthogonality relations (\ref{eq:projector_criteria}). The
derivation of Eq.\ (\ref{eq:eqcl_representation}) was given already in Appendix
\ref{app:eqcl_representation}. The only remaining case is therefore Eq.\
(\ref{eq:maximum_entropy}).

As noted in Sec.\ \ref{sec:equivalence_classes}, Eq.\ (\ref{eq:maximum_entropy})
is a consequence of Eqs.\ (\ref{eq:eqcl_representation}) and
(\ref{eq:increased_entropy}). The proof of Eq.\ (\ref{eq:increased_entropy}) in
the case of a partial reduction follows that given by von Neumann
\cite{vonNeumann1955secV4} for a total reduction.  To set the stage,
let us start by considering the transformation
\begin{equation}
\rho' = P_A \rho P_A ,
\end{equation}
in which $P_A$ is a projector that acts nontrivially only on the subsystems
in set $A$:
\begin{equation}
P_A = \tilde{P}_A \otimes 1_{B} .
\end{equation}
The density operator $\rho$ can be expanded as
\begin{equation}
\rho = \sum_{i,j} c_{ij} A_i \otimes B_j \qquad (c_{ij} \in \mathbb{C}) ,
\end{equation}
in which $A_i$ and $B_j$ are operators that act in $A$ and $B$,
respectively. Thus
\begin{equation}
\rho' = \sum_{i,j} c_{ij} (\tilde{P}_A A_i \tilde{P}_A) \otimes B_j .
\label{eq:rho_prime_P_K}
\end{equation}
The projector $\tilde{P}_A$ can be expanded as
\begin{equation}
\tilde{P}_A = \sum_{\mu = 1}^{d} \outprod{e_{\mu}}{e_{\mu}} ,
\end{equation}
in which $\{ \ket{e_{\mu}} \}$ is some orthonormal basis of eigenvectors of
$\tilde{P}_A$.  Equation (\ref{eq:rho_prime_P_K}) then becomes
\begin{equation}
\rho' = \sum_{\mu, \nu = 1}^{d} M_{\mu \nu}^A \otimes \tr_A (\rho M_{\mu
\nu}^{\dagger}) , \label{eq:rho_prime}
\end{equation}
in which
\begin{equation}
M_{\mu \nu} = M_{\mu \nu}^A \otimes 1_{B} , \qquad M_{\mu \nu}^A =
\outprod{e_{\mu}}{e_{\nu}} .
\end{equation}
For the special case of a rank-one projector (i.e., $d = 1$), 
we have $M_{11} = P_A$ and thus
\begin{equation}
\rho' = \tilde{P}_A \otimes \tr_A (\rho P_A) .
\end{equation}

This result can now be used to evaluate the successive transformations
\begin{equation}
\rho' = \sum_{\alpha} P_{A \alpha} \rho P_{A \alpha} = \sum_{\alpha}
\tilde{P}_{A \alpha} \otimes \tr_A (\rho P_{A \alpha})
\label{eq:transformation_P}
\end{equation}
and
\begin{equation}
\rho'' = \sum_{\beta} Q_{A \beta} \rho' Q_{A \beta} = \sum_{\beta} \tilde{Q}_{A
\beta} \otimes \tr_A (\rho' Q_{A \beta}) , \label{eq:transformation_Q}
\end{equation}
in which $\{ P_{A \alpha} \}$ and $\{ Q_{A \beta} \}$ are sets of rank-one
projectors. Following von Neumann \cite{vonNeumann1955secV4}, we can choose
these projectors to form complementary sets in each subspace defined by the
projectors $P_{i}$. The derivation given by von Neumann
\cite{vonNeumann1955secV4} then shows immediately that $\rho''$ is the same as
the reduced state $\hat{\rho}$ defined in Eq.\ 
(\ref{eq:rho_hat_new}). But Eqs.\ (\ref{eq:transformation_P}) and
(\ref{eq:transformation_Q}) are L\"uders reductions, which cannot decrease the
entropy $S(\rho)$ \cite{NielsenChuang2000}. Hence, the nondecreasing-entropy
property (\ref{eq:increased_entropy}) has been established, from which the
maximum-entropy property (\ref{eq:maximum_entropy}) follows.

% \bibliographystyle{abbrv}

%\bibliography{quantum}


\end{document}